\documentclass[review,authoryear]{elsarticle}

\usepackage{url}
\usepackage{amsmath}
\usepackage{maltese}
\author[1,2]{Joseph Caruana\corref{cor1}}
\ead{J.Migchielsen@elsevier.com}
\author[1]{Ryan Vella}
\author[1]{Daniel Spiteri}
\author[4]{Michael Nolle}
\author[3]{Sara Fenech}
\author[3]{Noel J.~Aquilina}

\cortext[cor1]{Corresponding author, joseph.caruana@um.edu.mt}

\address[1]{Department of Physics, University of Malta, Msida MSD 2080, Malta}
\address[2]{Institute of Space Sciences \& Astronomy, University of Malta, Msida MSD 2080, Malta}
\address[3]{Department of Geosciences, University of Malta, Msida MSD 2080, Malta}
\address[4]{Ambient Quality \& Waste Unit, Environment and Resources Authority, Malta}

\begin{document}

\begin{abstract}
Over the years, the Maltese Islands have seen a marked rise in the prevalence of artificial lighting at night.  The most evident type of light pollution arising from this evolution in anthropogenic night-time lighting is artificial skyglow via partial back-scattering in the atmosphere, leading to an increase in the Night Sky Brightness (NSB).  The importance of understanding and quantifying the geographical distribution of the NSB is underscored by the adverse impact of light pollution on various spheres, from astronomical observation to ecology and human health. For the first time, we present a detailed map of the NSB over the Maltese archipelago carried out with Unihedron Sky Quality Meters.  We show that the vast majority of the area of the Maltese Islands is heavily light polluted, with 87\% of the area registering a NSB $<$ 20.39~mag$_{\rm SQM}$/arcsec$^2$ (Bortle Class 5 or higher) and 37.3\% $<$ 19.09~mag$_{\rm SQM}$/arcsec$^2$ (Bortle Class 6 or higher), with the Milky Way being visible for only 12.8\% of the area (adopting a visibility threshold $>$ 20.4 - 21.29~mag$_{\rm SQM}$/arcsec$^2$; Bortle Class 4).  Coastal Dark Sky Heritage Areas on the island of Gozo retain generally darker skies than the rest of the islands, but light pollution originating further inland is encroaching upon and adversely affecting these sites.  The methodology presented in this study can be adopted for continued future studies in Malta as well as for other regions.
\end{abstract}

\begin{keyword}
night sky brightness - light pollution - site testing - techniques: photometric
\end{keyword}

\title{A Photometric Mapping of the Night Sky Brightness of the Maltese Islands}

\maketitle

\section{Introduction}

Artificial lighting is a cornerstone of modern society.  From functional illumination (e.g.~road lighting), to that which largely serves an ornamental purpose (e.g.~the lighting of public buildings and monuments), anthropogenic lighting is ubiquitous.  The increase in artificial illumination has brought with it a host of new challenges, from considerations of energy demand and efficiency to an urgency to understand artificial lighting's pollutant attributes.  In particular, numerous studies have revealed artificial lighting's disruptive impact upon ecology \citep[see, e.g.,][]{holker2010, gaston2013, gaston2014, bennie2016, manfrin2017}, and human health \citep[e.g.][]{chepesiuk2009, cho2015}, with both the prevalence and the spectrum of the light sources being important factors.

The advent of new technologies resulted in changes to the lighting sources that are used, perhaps the most significant of which was a transition from High Pressure Sodium (HPS) lamps to solid-state fixtures utilising Light Emitting Diodes (LEDs).  The Maltese Islands have also experienced this transition; road lighting and architectural lighting of churches and public buildings on one of the islands, Gozo, already consists almost exclusively of LED fixtures, and Malta is undergoing the same change.  As a result of the country's development over the years, the Night Sky Brightness (NSB) has perceptibly been increasing; in particular, since commencing continuous NSB monitoring in 2014, the conversion from HPS to LED fixtures resulted in a marked increase in the NSB (Nolle et al.~in prep.). However, no systematic country-wide assessment of light pollution has been published up to this point.  

The present study was motivated by a number of reasons.  Firstly, there was a need for a comprehensive survey of the NSB of the Maltese archipelago.  To date, the only archipelago-wide outlook has come from space-based images, e.g.~astronaut photography from the International Space Station and nighttime imagery from the Visible Infrared Imaging Radiometer Suite (VIIRS) on the NASA / NOAA Suomi National Polar orbiting Partnership (SNPP) satellite.  Whilst such images help to point out zones where there is upward light emission, they represent a different (and complementary) picture to the `view from the ground' as captured by ground-based measurements of the NSB, which is affected by partial backscattering from air particles, particulate matter (e.g.~aerosols), and other environmental factors.  

Similar NSB-mapping studies have been carried out in other regions, e.g.~Perth \citep{biggs2012}, Madrid \citep{zamorano2016}, Poland \citep{netzel2016}, and Eastern Austria \citep{posch2018}.  Such studies are, in turn, well placed to validate modelling approaches to the NSB.  In this work, we present one of the densest studies ever undertaken of the night sky brightness in a given region. Beside serving to map the NSB over the Maltese archipelago, this dataset is useful for comparison against theoretical studies that model and predict the NSB \citep[e.g.][which presents a world atlas of the artificial NSB]{falchi2016}.

The work was also aimed at looking at the potential impact of light pollution on Dark Sky Heritage Areas (DSHAs), local ecology and human health; we describe these aspects in further detail below.  Finally, the study was also intended to indicate the feasibility of the presently-adopted methodology for continued future monitoring.

\subsection{Dark Sky Heritage Areas (DHSAs)}
A 2006 local plan for the islands of Gozo and Comino approved by the Malta Environment and Planning Authority (MEPA) had designated a number of coastal zones on these two smaller islands as Dark Sky Heritage Areas.  The policy for such designated zones reads that ``reflective signs shall be employed to guide driving at night, whilst the installation of lighting which is not related to aerial or maritime navigation, shall be strongly discouraged'' \citep{localplan}.  The conservation of such sites is important for at least two reasons: offering a respite for nocturnal wildlife, and providing a place from where scientists and the general public alike can carry out astronomical observations.   With the proliferation of light pollution, many have lost the ability to observe a starry night sky, such that sections of the population (especially young people) have never seen the Milky Way. Beside eroding possibilities for scientific engagement and education, this unfortunate reality represents a loss of cultural heritage. One of the aims of the present work was to (i) investigate the present status of these DSHAs, particularly the impact upon them by light pollution originating from the rest of the islands, and (ii) possibly identify other zones (particularly on the island of Malta) that might be worthy of being assigned the same designation.  Quantifying the sky brightness in these zones was also required to help in identifying a suitable site for the setup of a modest astronomical observatory to be utilised both for research and public outreach purposes.  

\subsection{Ecology}
Within the local setting, one of the most pertinent ecological impacts of increased light pollution is upon certain seabird species.  The Maltese Islands host around 10\% of the global population of Yelkouan Shearwater (\emph{Puffinus yelkouan}), whose nesting is disturbed by light pollution \citep{birdlife2010}.  The problem can also affect Scopoli Shearwater (\emph{Calonectris diomedea}), Cory's Shearwater (\emph{Calonectris borealis}) and the European Storm-Petrel (\emph{Hydrobates pelagicus}), which are present in considerable numbers on the Maltese Islands; in particular, the latter constitutes the largest breeding colony in the Mediterranean \citep{raine2007}.  Beside the coastal zones that are of importance to such seabirds, the small size of the islands means that light pollution was expected to affect virtually every part of the country, since no region is sufficiently remote and isolated as to escape the reach and effects of nighttime illumination.  This could have repercussions for a number of other nocturnal creatures \citep[e.g.][]{brincat2014}. However, the effect upon biodiversity and ecosystems in a local context has not been comprehensively studied.  One of the aims of this study was to quantify the NSB in regions outside core urban areas, where such nocturnal species are more likely to be located.

\subsection{Human Health}
We were interested in identifying which urban zones exhibited the highest levels of NSB, with a view to establishing how widespread severe levels of light pollution were.

Numerous studies and reviews have pointed to a link between light pollution and problems of human health, including psychological wellbeing \citep{bedrosian2013}, breast cancer \citep[e.g.][]{haim2013, stevens2014, garcia-saenz2018}, and prostate cancer \citep[e.g.][]{haim2013, garcia-saenz2018} amongst others.  The hormone melatonin's effect upon cancer, in particular its role as an oncostatic, has been widely discussed in the literature \citep[e.g.][]{panzer1997, shiu1999, kanishi2000, anisimov2000, bondy2018}, and it has been suggested that exposure to nocturnal lighting can increase the risk of various cancers through the melatonin pathway \citep{schernhammer2004}.  Moreover, shorter-wavelength (blue) light has been pointed out as being a major disruptor to the human body's circadian rhythm, which receives input from intrinsically photosensitive retinal ganglion cells (ipRGCs), and it can suppress melatonin in humans \citep[e.g.][]{west2011}.  In particular, exposure for an equal amount of time to monochromatic 460~nm light resulted in twice the circadian phase delay and twice the melatonin suppression than exposure to 555~nm (i.e.~redder) light \citep{lockley2003}.  Another study that carried out a comparison between exposure to 5600~K and 2700~K light found that in addition to larger melatonin suppression by 5600~K light, the effect was significantly more pronounced in adolescents than adults \citep{nagare2019}.  Comparing the effect of 3000~K and 6200~K light amongst both children and adults,  \citet{lee2018} find that in children, blue-enriched LED light results in greater melatonin suppression.  Such findings may carry implications for LED fixtures that emit light with a correlated colour temperature $>3000$~K, such as street-lamps installed throughout all of Gozo and large parts of Malta.

Another manifestation of light pollution is glare, which causes discomfort, significantly reduces contrast sensitivity, and compromises the capability to distinguish colours. It is therefore detrimental to road safety and crime prevention.

\section{Methodology of Measurement}

\subsection{The Sky Quality Meter}
NSB measurements were carried out by means of a Unihedron Sky Quality Meter (model \emph{SQM-L}), which yields measurements in the astronomical standard of magnitudes per square arcsecond (mag$_{\rm SQM}$/arcsec$^2$).  The \emph{SQM-L} consists of a TAOS \emph{TSL237} light-to-frequency converter, whose response spans the range between 320 nm and 1050 nm, and is temperature compensated for the ultraviolet-to-visible range between 320 nm and 700 nm \citep{taos}.  The sensor is covered with an infrared-blocking filter (HOYA CM-500), and employs a lens that narrows down the angular sensitivity, for a Full Width Half Maximum (FWHM) of $\sim20^{\rm o}$ such that the sensitivity to a point source lying $\sim 19^{\rm o}$ off-axis is 10 times lower than for an on-axis point-source \citep{unihedron-sheet}.  As has been noted in the literature \citep{hanel2018}, this latter quality is important since it enables more consistent measurements in the presence of surrounding light sources. The manufacturer specifies a zero-point error of $\pm0.10$~mag$_{\rm SQM}$/arcsec$^2$ \citep{unihedron-sheet} and this was verified by, e.g.~\citet{puschnig2014}.  For further discussion on the technical aspects of the Sky Quality Meter, we refer the reader to \citet{cinzano2005}.  

The equipment setup for mobile data collection consisted of a tripod-mounted \emph{SQM-L}.  A digital level was used to accurately measure the zenith angle, and a physical compass was used to determine azimuth angles (for inclined measurements).  Such an approach for NSB 2D-mapping can be compared to other studies in the literature, e.g.~\citet{zamorano2016}. For our longterm, fixed-station readings in Gozo, the inclined \emph{SQM-L} was fitted with an anti-glare hood to block direct light (and thus avoid potential uncertainty) from a nearby streetlamp \citep{nolle2005}.  We found that without the hood in place, light from a single such streetlamp would affect measurements up to a distance of 60 m (or a zenith angle of 84$^{\rm o}$) away from the \emph{SQM-L}.

\subsection{The Geographic Information System}

The Geographic Information System (GIS) for this project was managed with the free, open-source software package \emph{QGIS}.  A 1 km$^2$-resolution grid was overlaid on top of a map of the Maltese Islands.  The choice of this resolution also finds support in \citet{bara2018}; adopting two separate tools, namely the luminance structure function and the Nyquist-Shannon spatial sampling theorem, this resolution was found be adequate for the determination of the zenithal NSB at any point to within $\pm 0.1$ mag$_{\rm V}$/arcsec$^2$. Our final grid-map contained a total of 347 cells.  Coordinates for data-acquisition sites were chosen to lie at the centre of each grid-cell.  However, in practice, it was not always possible to acquire data from these central coordinates (e.g.~for reasons of inaccessibility, or the presence of lighting fixtures whose emission would trespass onto the SQM-L).  In such instances, data were collected from the closest possible position to the cell-centre.

\subsection{Datasets}
\label{sec:datasets}
We acquired two independent datasets consisting of geographically-distributed zenithal NSB measurements.  Measurements for the first dataset were obtained on 34 nights between October 2017 and March 2018 (2017/18 dataset hereafter).  This dataset encompassed all of Malta, Gozo and Comino.  For consistency, data were collected only on moonless, cloudless nights.  Given the large-scale nature of the project, it was not feasible to record daily NSB values in each grid-cell.  However, a second dataset (utilising the same grid) was collected on the island of Malta on 14 nights between October and December 2018 (this dataset is called 2018/19 hereafter, since additional higher-resolution data for a specific zone were collected in 2019, as discussed in Section \ref{sec:resolution_section}); this enabled a comparison-check on the first dataset, in particular to assess any major variability in the data.  Data for both of these sets were collected using the same SQM-L.

In the case of the 2017/18 dataset, where possible, in addition to zenithal NSB readings, for grid-cells lying along the coast of the islands we also took NSB measurements at zenith angles of $30^{\rm o}$ and $60^{\rm o}$, facing both away from the coast (i.e.~towards the sea) and inland.  These data were collected as they are useful for: 

\begin{enumerate}
\item Assessing coastal zones (where all DSHAs are located) which should offer relatively darker skies facing the sea (as opposed to facing inland), in so doing establishing the impact that nearby zones have on some of these relatively darker coastal sites. \\
\item Gaining insight into what bird species would perceive as they flew towards / away from land.
\end{enumerate}

Further to the above two datasets, over a five-year period we continuously recorded the NSB from a fixed location in the West of Gozo (San Lawrenz) using two SQM-LU-DLs (which have the same characteristics, e.g.~FOV and response, as the SQM-L) - one pointing to the zenith, and another inclined at a zenith angle of 45$^{\rm o}$ and azimuth angle of 106$^{\rm o}$.  This azimuth is the approximate direction of the strongest horizontal sky brightness, as the field of view of this inclined meter overlooks the densely populated areas of Gozo and Malta. Both meters are configured to take readings in five-minute intervals. This dataset allows us to look at variations of the NSB due to varying meteorological conditions as well as long-term trends.  

Data were collected over 1803 nights, recording readings once the sky became darker than 12 mag$_{\rm SQM}$/arcsec$^2$ for a total of 239,079 readings. In analysing the NSB trend over this 5-year period, we retained data collected between 21:00 to 03:00, such that each night consists of 73 individual readings for a total of 131,619 measurements; of these, we retained 19\% after filtering the data for clear and moonless periods. For the station in question, the data were filtered to keep zenithal readings $>20$~mag$_{\rm SQM}$/arcsec$^2$ during the HPS period and $>19.5$~mag$_{\rm SQM}$/arcsec$^2$ after street lamps were retrofitted with LED fixtures. We required a difference between the zenithal and inclined readings of $<1.3$ mag$_{\rm SQM}$/arcsec$^2$. These are empirical values and apply for weather conditions of homogeneous Cirrus cloud coverage.  Furthermore, since in the case of clear skies we do not expect quick changes in the NSB, we required that any change between two consecutive measurements (5 mins apart) for both meters individually as well as their difference to be less than 0.1~mag$_{\rm SQM}$/arcsec$^2$. Nolle et al.~(in prep.) will provide a more detailed description of the setup, data screening, quality assurance and results.

\section{Results and Analysis}

\subsection{NSB Map of the Maltese Islands}

\begin{figure}
  \includegraphics[width=\linewidth]{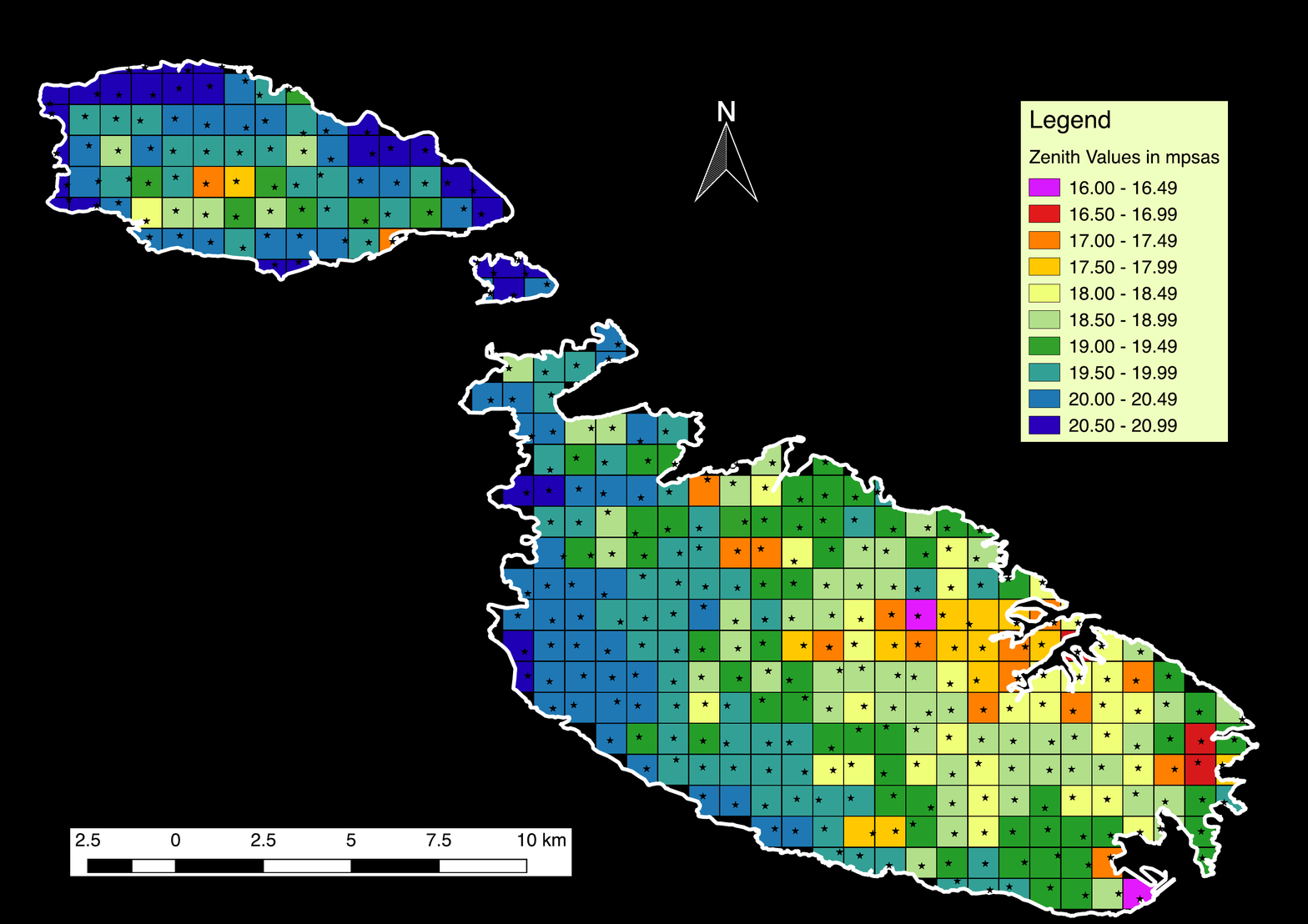}
  \caption{A 2D grid-map of the zenithal NSB in mag$_{\rm SQM}$/arcsec$^2$ over the Maltese archipelago.  Stars within each 1 km$^2$ cell mark the exact position where measurements were obtained.}
  \label{fig:data1718}
\end{figure}

The 2017/18 dataset allowed us to build a two-dimensional grid-map of the NSB over the entire Maltese archipelago, as shown in Fig.~\ref{fig:data1718}.  

Taking the natural sky brightness to be 22~mag$_{\rm SQM}$/arcsec$^2$ (corresponding to 174~$\mu$cd/m$^2$, as in \citet{falchi2016}; \citet{plauchu-frayn2017} measure 21.88~mag$_{\rm{SQM}}$/arcsec$^2$), areas exhibiting a NSB $> 20.99$~mag$_{\rm SQM}$/arcsec$^2$ and $> 16$~mag$_{\rm SQM}$/arcsec$^2$ are  respectively $2.54\times$ and $251\times$ brighter than the natural sky brightness.

\begin{figure*}
  \includegraphics[width=\linewidth]{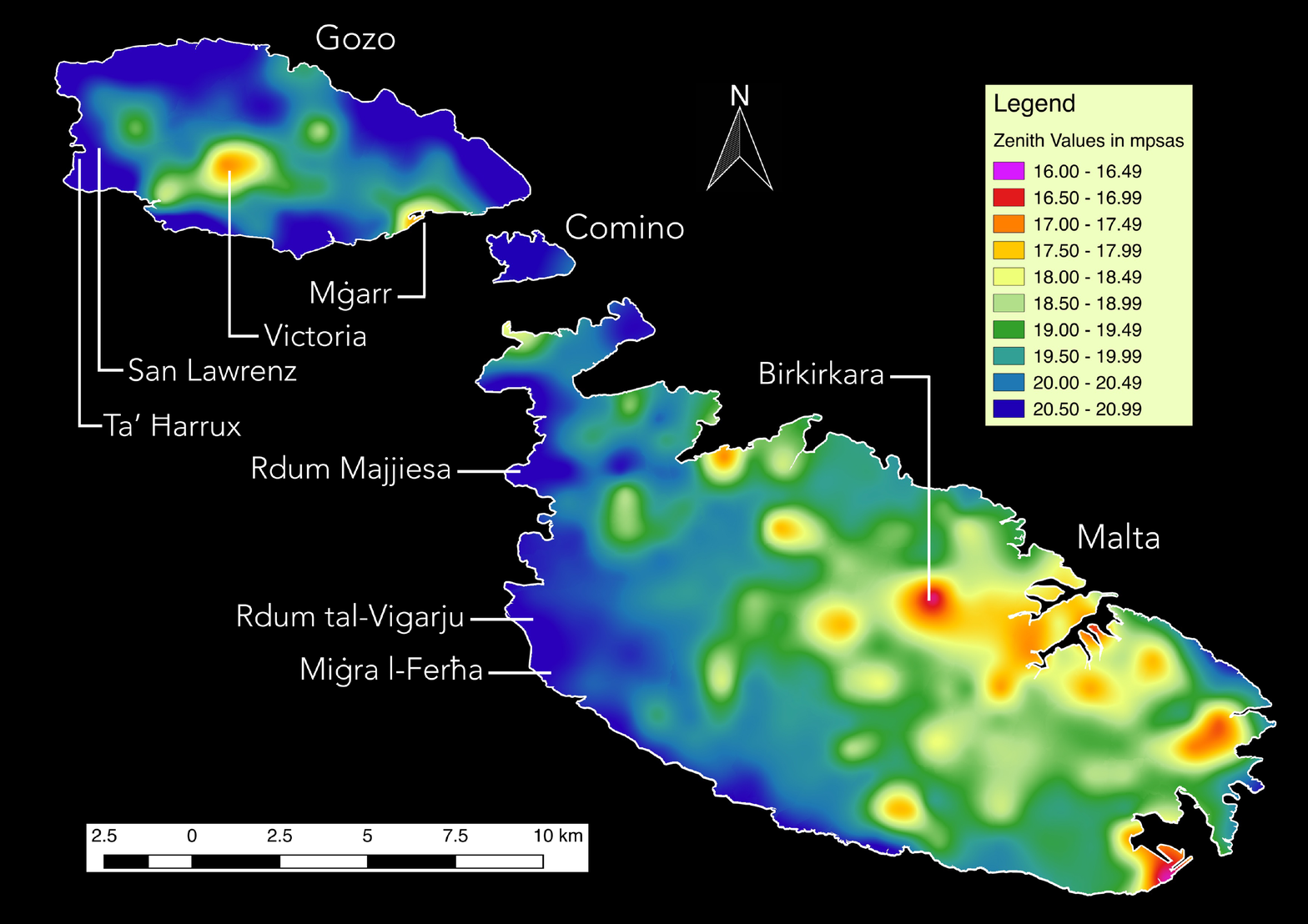}
  \caption{A thin-plate-spline interpolated 2D-map of the NSB over the Maltese archipelago based on the 2D grid-map of Fig.~\ref{fig:data1718}.}
  \label{fig:tps1718}
\end{figure*}

From our grid-map, we created a thin-plate-spline interpolated map (TPS-map hereafter), shown in Fig.~\ref{fig:tps1718}.  This interpolation method derives its name from the physical analogy of a thin metal sheet that is bent to fit a finite set of points, whereby one minimises the non-negative integral quadratic variation (or `bending energy'), $I_{f(x,y)}$, that is given by \citet{bookstein1989}:

\input{amssym.tex}
\begin{equation}
I_{f(x,y)}=\iint_{{\Bbb R}^2} \left(\frac{\partial^2 f}{\partial x^2}\right)^2+2\left(\frac{\partial^2 f}{\partial x \partial y}\right)^2+\left(\frac{\partial^2 f}{\partial y^2}\right)^2 dx dy
\end{equation}

We considered a number of other interpolation methods that included nearest-neighbour, cubic-spline, ordinary kriging and inverse-distance weighting. In nearest-neighbour interpolation, for any point $S_0$ at which the value of the function, $f(S_0)$, is to be determined, the latter will always assume the value of the nearest data-point.  This effectively restricts the function's co-domain exclusively to the set of known values, which is unsuitable for the purpose at hand.  Cubic-spline interpolation does not represent extreme values appropriately, with NSB maxima being smoothened down to lower values.  Ordinary kriging yields an extremely smoothened representation of the data, washing out all detail.  Inverse-distance weighting represents the best option in comparison to the previous three methods, and is useful for resolving `clustered' NSB peaks that are spatially very close to each other, yielding a more distinct representation of the data than the one resulting from thin-plate-spline interpolation.  However, it occasionally washes out extreme values.  Moreover, within the practical context of this study, given the small scales over which its spatial feature separation applies, minor variations in NSB are likely to dominate.  Thin-plate-spline does not alter the pattern or shape, and does not suppress peak values.  Given the above considerations, thin-plate-spline interpolation was deemed the best option to visually represent the data.

\subsection{Inclined measurements}

\begin{figure}
  \includegraphics[width=\linewidth]{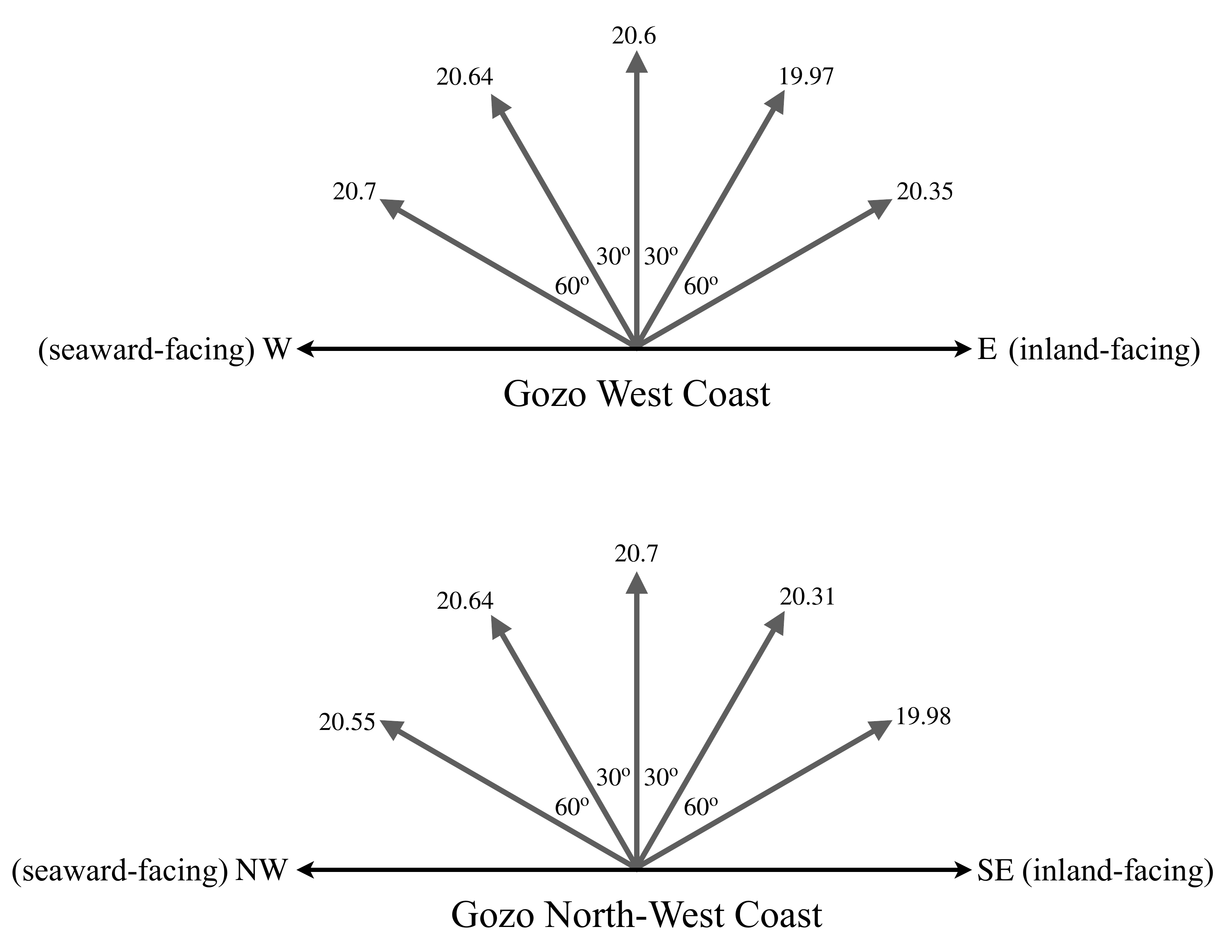}
  \caption{Two examples of data collected from DSHAs in the West and North-West coast of Gozo that show how these coastal zones are affected by light pollution originating further inland, resulting in a gradient in the measured NSB, with the sky being darker facing seaward (as opposed to inland).}
  \label{fig:gradient_plots}
\end{figure}

As mentioned in Sec.~\ref{sec:datasets}, for the 2017/18 dataset, in coastal grid-cells we also obtained measurements with a zenith angle of 30$^{\rm o}$ and 60$^{\rm o}$, facing both towards and away from land.  We first determined the centroid of the polygon maps of Malta and Gozo in \emph{QGIS}.  Then we divided each map into eight 45$^{\rm{o}}$ segments, each segment being delineated on either side by one of the cardinal or intercardinal directions. For a given coastal grid-cell, the azimuth angle of the meter depended on which compass point most of the cell's area subscribed to. As an example, considering a coastal cell lying between the NW and N directions, we would first draw a straight line from the map centroid to the cell centre.  Next, we would determine whether most of the cell's area was closest to the NW or the N direction, and finally we would choose the meter's azimuthal direction to be one of these two accordingly. Such data enabled us to assess the impact of light pollution originating further inland upon the (coastal) DSHAs, with two examples shown in Fig.~\ref{fig:gradient_plots}.  As expected, the data revealed a difference in the NSB at these sites depending on whether one faced inland or seaward, with the sky being generally significantly darker looking away from land and out towards the sea.  Nevertheless, there were cases where this was not the case due to light spilling over from other zones. For example, the South-East harbour in M\malteseg arr (Gozo) suffered from significant sky brightening overlooking the sea due to light pollution both from harbour lights and spillage from Malta.  Even the remotest parts of the islands, such as the West and North-West coast of Gozo, were not immune to a gradient in the NSB, registering a difference of 0.73 mag$_{\rm SQM}$/arcsec$^2$.  This finding clearly underlines the widespread impact of light pollution arising from urban zones.  This points to a need to establish a perimeter and $\approx 0.5$ km buffer zones around DSHAs with development policies that strictly limit the luminous flux and temporal use, and impose further reduction in blue light emission of outdoor lighting.   The latter could consist of Phosphor-Converted LEDs (PC-LEDs) which emit light with a correlated colour temperature of 1700-2500~K.  Taking into account the diminished colour-discrimination of human vision in low light, the estimated general colour-rendering index of these lights lies between that of white phosphor converted LEDs and HPS lamps \citep{zabiliute2014}.  Such PC-LEDs and LED 2700~K-filtered lamps are both astronomy-friendly and have a lower potential impact upon melatonin suppression than HPS lamps \citep{aube2013}.

\subsection{Sky Brightness Classification}

The naked eye observability of an astronomical object of a given magnitude depends on the brightness of the sky background.  A 9-point numeric scale that descriptively classifies the quality of the night sky in terms of the visibility of various astronomical objects was introduced in 2001 \citep{bortle2001}.  Having come to be known as `The Bortle Scale' (after its author), it has seen widespread use, since a classification according to this scale can immediately convey a general idea of the visibility of various astronomical objects, such as the Milky Way. The Bortle scale can be approximately related to values of NSB in magnitudes per square arcsecond\footnote{See: http://www.darkskiesawareness.org/img/sky-brightness-nomogram.gif}; whilst this conversion is approximate, it is nevertheless useful to categorise the sky brightness in a number of such brightness bins, and assess the percentage of land-area that falls within each sky-brightness category.  The results for the Maltese Islands are presented in Table \ref{table:bortle}.

In over 87\% of the area of the Maltese Islands, we record a NSB \sloppy $<20.39$~mag$_{\rm SQM}$/arcsec$^2$, and for 37.3\% of the area we record a NSB $<19.09$~mag$_{\rm SQM}$/arcsec$^2$.  As expected, the most light polluted of the Islands is Malta, with 96\% of the area registering a NSB $<20.39$~mag$_{\rm SQM}$/arcsec$^2$ and 46\% $<19.09$~mag$_{\rm SQM}$/arcsec$^2$.  The brightest recorded value in Malta was 16.36~mag$_{\rm SQM}$/arcsec$^2$ in Birkirkara, whereas the darkest was 20.61~mag$_{\rm SQM}$/arcsec$^2$ in Rdum tal-Vigarju (cliffs off Ba\malteseh rija, limits of Rabat).

Overall, Gozo remains darker than Malta, registering a NSB \sloppy $<20.39$~mag$_{\rm SQM}$/arcsec$^2$ in 62.8\% of the area and $<19.09$~mag$_{\rm SQM}$/arcsec$^2$ in 11.5\% of the area.  The brightest recorded value of 17.24~mag$_{\rm SQM}$/arcsec$^2$ was recorded in Victoria, and the darkest value of 20.71~mag$_{\rm SQM}$/arcsec$^2$ was registered at Ta' \malteseH arrux (off Dwejra, limits of San Lawrenz). The Milky Way does not have a bearing on these results. In general, the sky was too bright for the Winter Milky Way's contribution to be significant. For darker ($>20$ mag$_{\rm SQM}$/arcsec$^2$) skies, it was not crossing the zenith when the measurement was acquired.

Comparing our results from the 2017/18 dataset (covering all three main islands) to \citet{falchi2016}, we find that a larger percentage of the area exhibits upper-end NSB values; in our dataset, $\approx$33\% of the area exhibits a NSB $>3000$~$\mu$cd$/m^2$ ($<18.89$ mag$_{\rm SQM}$/arcsec$^2$) whereas \citet{falchi2016} report $\approx$17\%.

\begin{table*}
\small
\begin{center}
 \begin{tabular}{|| c || c || c | c | c | c | c ||} 
 \hline
Bortle & mag$_{\rm{SQM}}$/ & Malta & Malta & Gozo & Comino & All \\ [0.5ex] 
Scale & arcsec$^2$ & (\%) & (\%) & (\%) & (\%) & (\%) \\
 & & 2017/18 & 2018/19 & 2017/18 & 2017/18 & 2017/18 \\
 \hline\hline 
Class 1 & 21.70 - 22.00 & 0 & 0 & 0 & 0 & 0 \\ 
Class 2 & 21.50 - 21.69 & 0 & 0 & 0 & 0 & 0 \\
Class 3 & 21.30 - 21.49 & 0 & 0 & 0 & 0 & 0 \\
Class 4 & 20.40 - 21.29 & 3.9 & 1.6 & 37.2 & 83.3 & 12.8 \\
Class 5 & 19.10 - 20.39 & 50.2 & 48.6 & 51.3 & 16.7 & 49.9 \\
Class 6, 7 & 18.00 - 19.09 & 34.0 & 35.8 & 7.7 & 0 & 27.4 \\
Class 8, 9 & $<$17.99 & 12.0 & 14.0 & 3.8 & 0 & 9.9 \\ [1ex] 
 \hline 
\end{tabular}
\caption{This table shows the percentage of land surface area falling within given brightness ranges, with the ranges being related to classes of the Bortle scale.  Archipelago-wide data is available only for 2017/18, so the values presented here base upon that dataset, but for the island of Malta percentages are also available for the 2018/19 dataset.}
\label{table:bortle}
\end{center}
\end{table*}

\subsection{Resolution considerations}
\label{sec:resolution_section}

\begin{figure}
  \includegraphics[width=\linewidth]{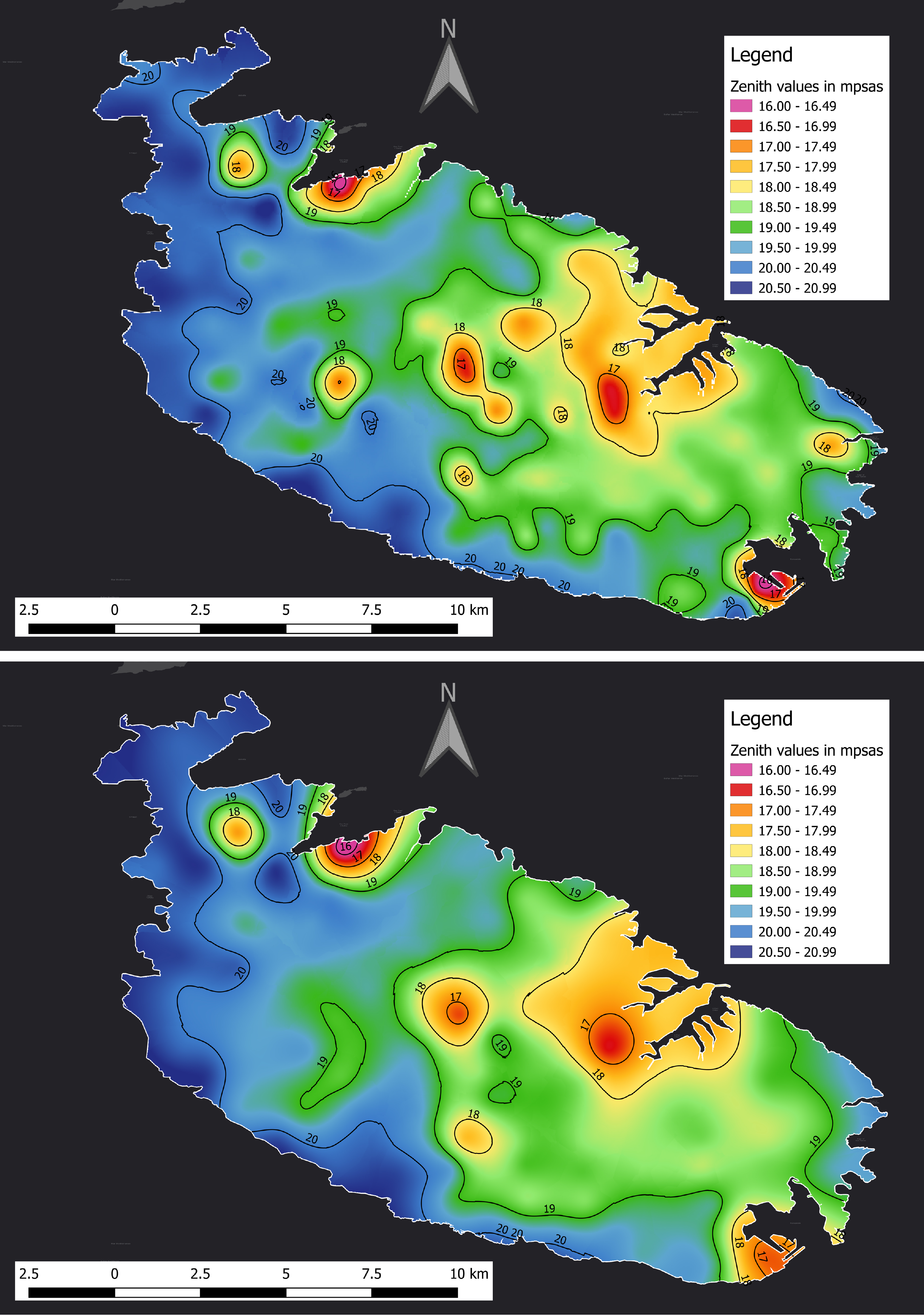}
  \caption{The top panel shows TPS-interpolated NSB 2018/19 data produced from the full 257-cell grid, whereas the bottom panel shows the same but with every other cell skipped, effectively utilising half the total number of cells and thereby reducing the resolution by half. The latter being less time-intensive to produce by virtue of the smaller number of cell-measurements required, this setup could be effectively employed for annual monitoring.}
  \label{fig:fullvshalf}
\end{figure}

The data-collection approach presented in this paper is very time-intensive, as it entails data acquisition from 347 individual cells.  Combined with the requirement that the data are collected on moonless, cloudless nights, this makes the annual monitoring of the NSB a challenging prospect from a timing perspective.  In view of this, we carried out a qualitative comparison of our 1 km$^2$-resolution map with one of lower resolution.  Considering the map of Malta from the 2018/19 dataset, we produced a second map for which we skipped every other data-cell, effectively reducing the number of contributing cells by half.  The result of this exercise is shown in Fig.~\ref{fig:fullvshalf}.

It is apparent from this comparison that whilst (as expected) a small number of hotspots are missed in the lower-resolution version (and a level of fine-structure detail is lost) by virtue of the alternating-cell approach, the overall NSB pattern is adequately captured. 

We were also interested in considering, as a case example, a given bright zone at a higher resolution of 100 m $\times$ 100 m, with a view to relating the brightness recorded in the coarser 1 km$^2$ grid-cell to light emanating from specific resolved structures (e.g.~roads).  In this instance, we found that unregulated shop and showroom lighting, badly designed exterior household lighting fixtures, and unshielded lighting ornaments were the main contributors to elevated NSB levels (see also Section \ref{sec:sources}), with their contribution clearly resolvable in this higher-resolution analysis.

Via these exercises, we established that annual monitoring of the NSB over the entire archipelago can be adequately carried out using half the resolution of our present 1 km$^2$-cell grid. On the other hand, a 100 m $\times$ 100 m grid is appropriate for detailed zonal studies aimed at resolving, and thereby identifying, sources of light pollution.

\section{Discussion}

\begin{figure}
  \includegraphics[width=\linewidth]{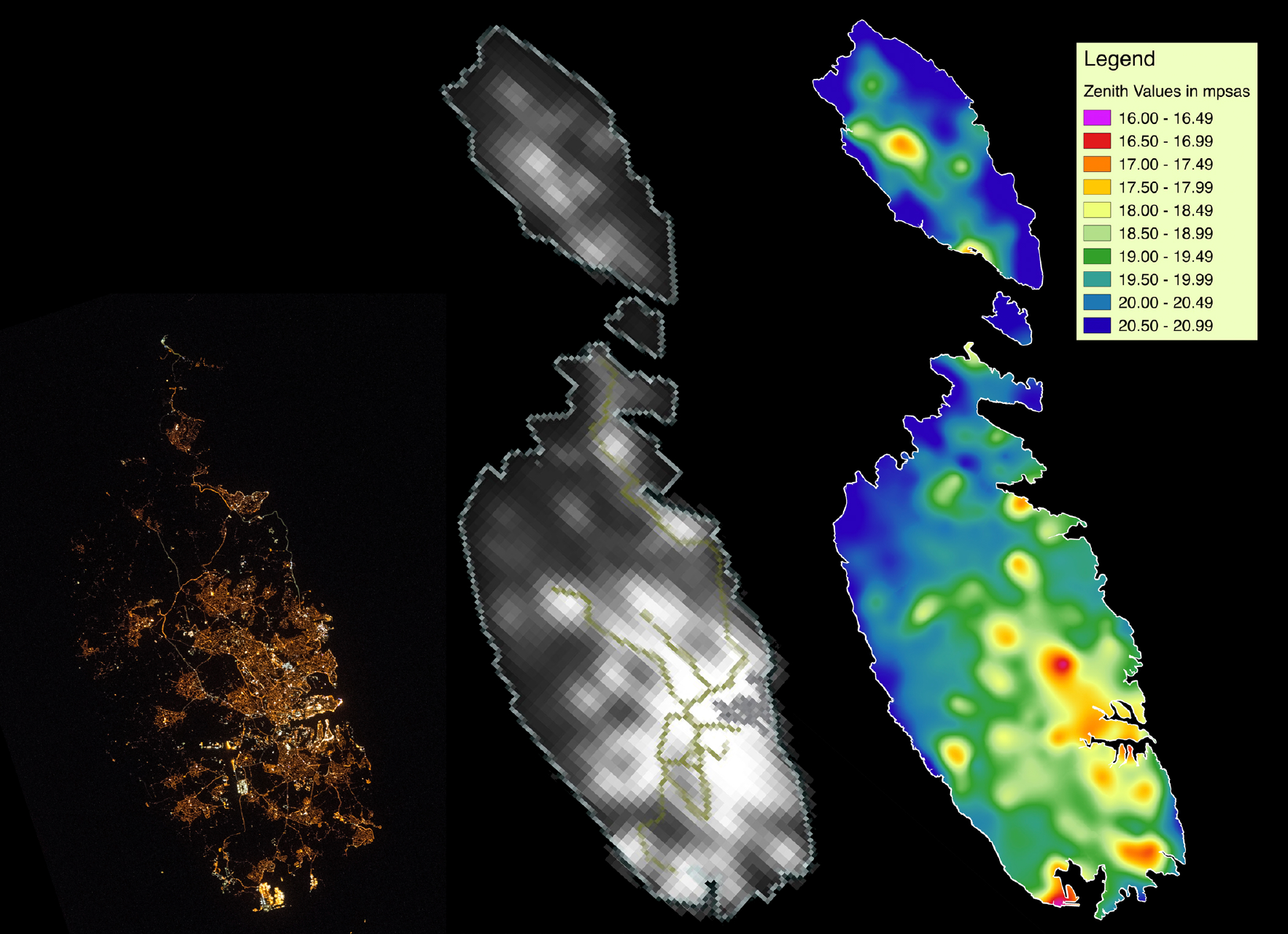}
  \caption{The image on the left shows Malta at night as viewed from the International Space Station.  The photograph (ISS047-E-55285) was captured on the 8th of April 2016 at 22:15:27 U.T. with a Nikon D4 (employing a 36$\times$23.9 mm 16.2 megapixel CMOS sensor) at a focal length of 400~mm, aperture of $f/2.8$, shutter speed of $1/40$~s and ISO 10,000.  The panel in the centre shows a night-time image obtained on the 22nd of November 2018 by the Visible Infrared Imaging Radiometer Suite (VIIRS) on the NASA/NOAA Suomi National Polar orbiting Partnership (SNPP) satellite.  The image on the right is a rotated version of our Fig.~\ref{fig:tps1718} based on our 2017/18 dataset, showing our ground-based NSB measurements. \textit{Left image courtesy of the Earth Science and Remote Sensing Unit, NASA Johnson Space Centre.}}
  \label{fig:spacevsground}
\end{figure}

A visual comparison of our ground-based NSB map with space-based imagery, specifically from the Visible Infrared Imaging Radiometer Suite (VIIRS) on the NASA/NOAA Suomi National Polar orbiting Partnership (SNPP) satellite, shows general agreement in the emission pattern. Furthermore, comparison with photography obtained from the International Space Station (ISS) reveals a correlation between elevated levels of NSB and those areas that exhibit a high concentration of artificial lighting emitting skyward (see Fig. \ref{fig:spacevsground}).

\begin{figure}
  \includegraphics[width=\linewidth]{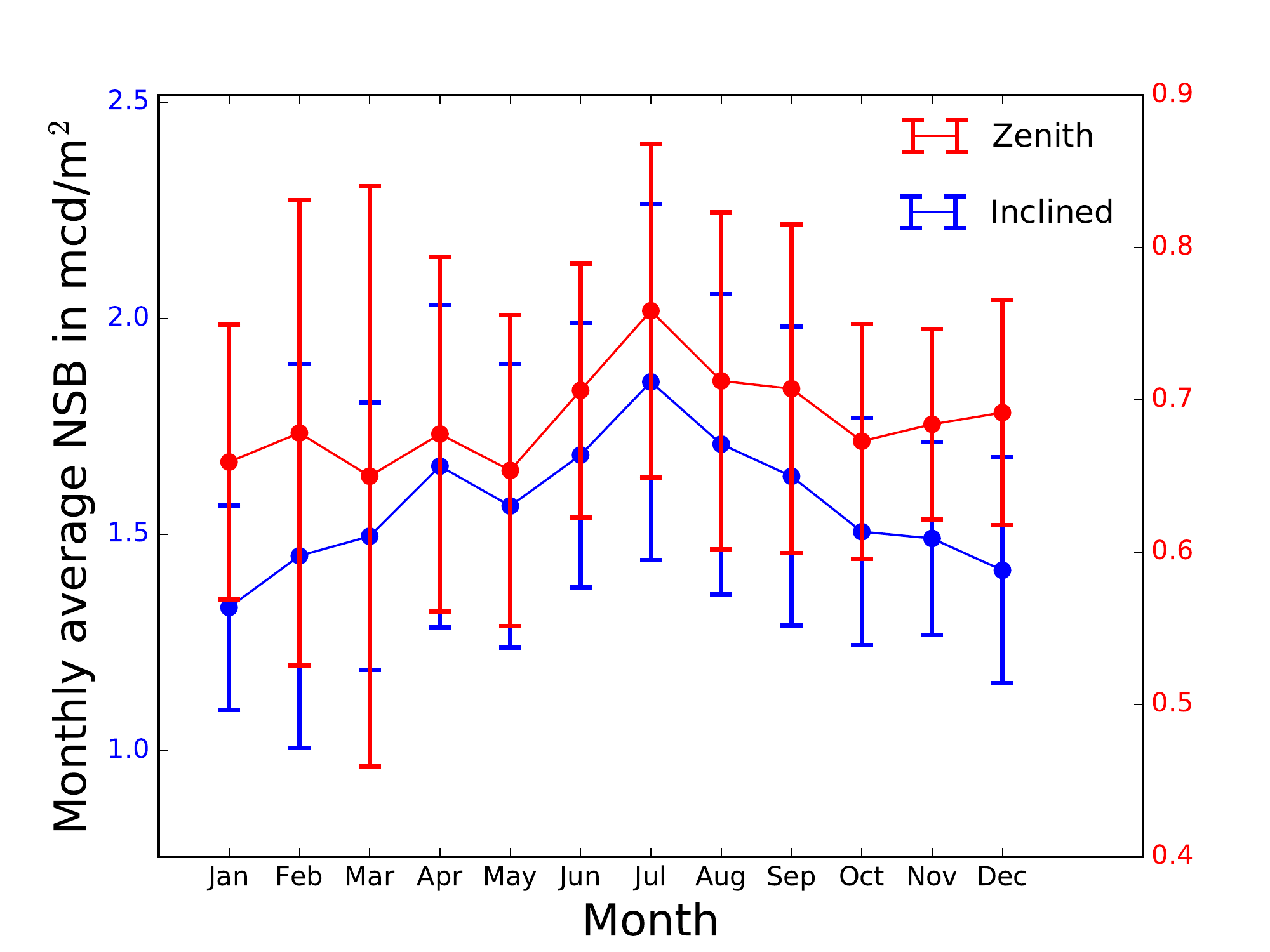}
  \caption{This figure shows the longterm NSB measured between 2014 and 2019 from a fixed site in Gozo.  Each datapoint represents the mean for a given month over this five-year period.  One can note a seasonal pattern, with the NSB increasing over the summer months.}
  \label{fig:seasonalvariability}
\end{figure}

The results presented in our maps can be considered to represent an optimistic scenario for two main reasons.  Firstly, they present measurements at the zenith, which is generally the sky's darkest region. Secondly, measurements were obtained during the winter months, when sky transparency is generally better.  Conversely, if the same exercise were carried out over the summer months, we expect to find higher NSB values, which is supported by Fig.~\ref{fig:seasonalvariability} which shows the monthly mean and standard deviation of the NSB measured with the inclined meter over five years from the single fixed-site station in San Lawrenz. The data from the meter were converted from mag$_{\rm SQM}$/arcsec$^2$ to mcd/m$^2$ in order to better visualise this seasonal variation. For consistency, we only used data from moonless and cloud-free periods. The diurnal time window was further restricted to between 21:00 and 3:00 CET in order to avoid assigning more statistical weight to the longer winter months. (Moreover, early evenings and late mornings are often influenced by particular meteorological effects and strong temporary sources, e.g.~the flood lighting of a local sports ground in the evening.) Overall, July experiences the brightest sky with 1.85 mcd/m$^2$ (19.42~mag$_{\rm SQM}$/arcsec$^{2}$) while January has on average the darkest night sky with 1.33 mcd/m$^2$ (19.77~mag$_{\rm SQM}$/arcsec$^{2}$).  The Environment and Resources Authority (ERA) operates five stations that collect particulate matter (PM) data.  Four of these are located in Malta (Attard, Kordin, Msida and \.{Z}ejtun), and one in Gozo (G\malteseh arb).  From an analysis of this PM data spanning ten years (2008-2018), we find no significant seasonal variability of particulate matter (PM$_{2.5}$, PM$_{10}$ and PM$_{\rm coarse}$) concentration throughout the year, so this does not seem to be a factor contributing to the seasonal variability that we observe in the NSB.

Higher NSB values would also be expected in overcast conditions.  \citet{ribas2016} and \citet{posch2018} find evidence that the effect of cloud cover is different for urban vs.~rural areas, with the former brightening further and the latter exhibiting darker skies, the proposed explanation being that for rural areas, clouds block contributions from the Milky Way, Zodiacal light and stars (which would otherwise generate brighter values).  However, all rural areas of the Maltese Islands are brighter than those considered in \citet{ribas2016}, with the darkest recorded value being 20.71~mag$_{\rm SQM}$/arcsec$^{2}$, so this effect would probably not come into play in the case of the Maltese Islands.

Moreover, the NSB as perceived by scotopic vision (i.e. the rod-dominated low-light human vision) is higher than that measured by SQMs, meaning that the SQM readings do not represent the scotopic magnitude, and therefore underestimate the brightness level as perceived by human vision \citep{sanchez2017}.  The combination of all of the above factors means that, overall, the results presented in this study present an optimistic scenario.

Considering the mean nightly zenithal NSB for a single year (2016), and including all data, i.e.~even those acquired during moonlit and cloudy conditions, we find that the frequency distribution is spread between 16 and 20.8 mag$_{\rm SQM}$/arcsec$^2$, and peaks at a NSB value of 20 - 20.2 mag$_{\rm SQM}$/arcsec$^2$. This peak corresponds to clear (moonless and cloudless) nights. For comparison, data acquired during the same year in Eastern Austria by \citet{posch2018} at a station located in Z\"{o}blboden finds this peak to be located at 21.6 mag$_{\rm SQM}$/arcsec$^2$, whereas measurements acquired in 2013 at Leopold-Figl Observatory in Austria by \citet{puschnig2014} yield a peak at $\approx$ 20.7 mag$_{\rm SQM}$/arcsec$^2$.  A study by \cite{bertolo2019} in the Veneto region of Italy reports data recorded in 2018 and finds peak values ranging between $\approx18.3$ mag$_{\rm SQM}$/arcsec$^2$ (Padova) and $\approx21.4$ mag$_{\rm SQM}$/arcsec$^2$ (Passo Valles).  Again, our peak represents an optimistic scenario since the station from which we recorded data continuously is located at one of the darker sites on the islands.

\subsection{Sources of Upward-Emitting Light}
\label{sec:sources}

In carrying out our measurements across the islands, we identified sources that visibly contributed a large amount of upward-directed light.  Some sources are permanent, in that they are switched on all night all year round, whereas others are transitory. Moreover, whilst certain cases do entail required lighting, the vast majority of examples involve illumination that is purely ornamental.

\begin{table}
\begin{center}
 \begin{tabular}{|| l || c || c ||} 
 \hline
 Source & Permanent & All night \\ [0.5ex] 
 \hline\hline 
Billboards & Yes & Yes \\
Churches & Yes & Yes \\
Football Grounds & No & No \\ 
Harbours & Yes & Yes \\
Monuments & Yes & Yes \\ 
Playgrounds & Yes & Yes \\
Public Squares & Yes & Yes \\
Shops/Showrooms & Yes & Yes \\ [1ex]
 \hline 
\end{tabular}
\caption{This table lists the main sources of badly - designed lighting fixtures contributing upward-emitted light that were identified in the course of this study.  The `Permanent' column denotes whether a source is switched on every night, and `All night' whether the light in question is switched on all night.}
\label{table:sources}
\end{center}
\end{table}

The main permanent culprits were found to be public monuments, churches, shops, showrooms, billboards, public squares, and playgrounds, all of which could employ smart lighting and be subject to specific curfews (see Table \ref{table:sources}). 

We also find evidence that recently-installed, badly-implemented road lighting has resulted in excessive night sky brightening.  Specifically, LED fixtures installed in a 1.4 km road (Mellie\malteseh a bypass) in the time between the acquisition of our 2017/18 and 2018/19 datasets increased the local NSB by more than two whole astronomical magnitudes (19.86 $\rightarrow$ 17.24~mag$_{\rm SQM}$/arcsec$^2$), effectively representing an 11$\times$ increase in sky brightness; this inordinate increase in the NSB may be attributed to excessive luminance levels of blue-rich (4000 K) LED fixtures that were installed in this stretch of road. Better monitoring of DSHAs is also required; for example in Dwejra, on-site artificial lighting resulted in an increase of nearly a whole astronomical magnitude (20.61 $\rightarrow$ 19.7 ~mag$_{\rm SQM}$/arcsec$^2$).

\section{Conclusions}

Measurements of the NSB via SQM-L meters in a regular 1 km$^2$ grid were used to create a 2D-map of the NSB; halving the resolution to 2 km$^2$ yields a map that still adequately captures the NSB distribution, and this method can be used for annual monitoring of the NSB over the Maltese Islands.  An interpolated version of the grid map was created for visualisation purposes, with thin-plate-spline interpolation found to be the most adequate amongst other tested interpolation algorithms, which included nearest-neighbour, cubic-spline, ordinary kriging and inverse-distance weighting.  Whilst we have so far carried out all our measurements on land, we intend to expand our studies to carry out offshore measurements of the NSB, adopting a method similar to that described in \citet{ges2018}; this would lend further applications of use to the present datasets. We also plan to add a second continuous-monitoring station. This would be placed at an urban site to complement the data acquired from our present single station situated in a darker zone, with a view to assessing whether (depending on the adoption of better lighting practices) an improvement might be registered in the prevailing situation in urban areas.

\subsection{Main findings}

\begin{enumerate}
\item The vast majority of the area of the Maltese Islands is heavily light polluted, with over 87\% of the area registering a NSB $<20.39$~mag$_{\rm SQM}$/arcsec$^2$ (Bortle Class 5 or higher) and 37.3\% $<19.09$~mag$_{\rm SQM}$/arcsec$^2$  (Bortle Class 6 or higher).  As expected, the most light polluted of the islands is Malta, with 96\% of the area registering a NSB $<20.39$~mag$_{\rm SQM}$/arcsec$^2$ (Bortle Class 5 or higher) and 46\% $<19.09$~mag$_{\rm SQM}$/arcsec$^2$ (Bortle class 6 or higher).  There is not a single location on the Islands exhibiting a NSB $>$ 21.3~mag$_{\rm SQM}$/arcsec$^2$ (Bortle Scale 3 or better), effectively meaning that there is no zone on any of the islands of Malta, Gozo or Comino that exhibits pristine dark skies. Gozo remains darker overall, aided both by its physical distance from Malta and by the fact that a number of areas are designated as DSHAs.  These results underline the requirement for a nationwide strategy on light pollution.
\item We find mild seasonal variability in the NSB, with summer months exhibiting higher (monthly-averaged) NSB values.
\item Particulate matter (PM$_{2.5}$, PM$_{10}$ and PM$_{\rm coarse}$) concentration does not exhibit seasonal variation, and therefore is not likely to be one of the factors contributing to the observed seasonal variability in the NSB.
\item Considering the Milky Way to be clearly visible only for Bortle Class 4 (20.4 - 21.3~mag$_{\rm SQM}$/arcsec$^2$) or better, the Milky Way is visible for only 12.8\% of the Maltese Islands (2017/18 dataset).  \citet{falchi2016} adopt a threshold of 20.0 - 20.6~mag$_{\rm SQM}$/arcsec$^2$ for Milky Way visibility, on which basis it is stated that the Milky Way is visible for only 11\% of the area of the islands. Adopting the midpoint of this range as the threshold ($>20.3$~mag$_{\rm SQM}$/arcsec$^2$), we find that the Milky Way is visible for 13.5\% of the islands, which is in good agreement with the result of \citet{falchi2016}.  (Adopting thresholds at either extreme of this range, namely $>20.6$~mag$_{\rm SQM}$/arcsec$^2$ and $>20$~mag$_{\rm SQM}$/arcsec$^2$, results in 6.3\% and 25.9\% respectively.)
\item Coastal DSHAs retain generally darker skies than the rest of the islands within the Maltese archipelago.  However, light pollution originating from outside these designated areas is having an adverse effect on them, necessitating proper light pollution mitigation measures beyond the DSHAs themselves.
\item In Gozo, the highest levels of NSB are registered in Victoria and G\malteseh ajnsielem. Addressing these two zones with a view to limiting unnecessary and/or exceedingly bright night-long lighting would help mitigate the growing issue of light pollution spreading to Dark Sky Heritage Areas.
\item There is scope for designating the site of Rdum Majjiesa (close to Il-Majjistral Nature \& History Park) and the area stretching between Rdum tal-Vigarju and Mi\.{g}ra l-Fer\malteseh a as DSHAs, on the basis that the recorded zenithal NSB in these zones ($20.54$~mag$_{\rm SQM}$/arcsec$^2$  and $20.55-20.61$~mag$_{\rm SQM}$/arcsec$^2$ respectively) is comparable to that in current DSHAs in Gozo and Comino.
\item Main sources of light pollution include public monuments, churches, football grounds, harbours, showrooms, and excessive street lighting, in many cases involving illumination from flood-lighting and $>3000$~K LED fixtures.
\end{enumerate}

\section{Acknowledgements}

We thank the Environment and Resources Authority for making available PM$_{2.5}$ and PM$_{10}$ data for analysis.

\appendix

\renewcommand\thefigure{\thesection.\arabic{figure}}

\section{Supplementary Material}

\setcounter{figure}{0}

\begin{figure}
  \includegraphics[width=\linewidth]{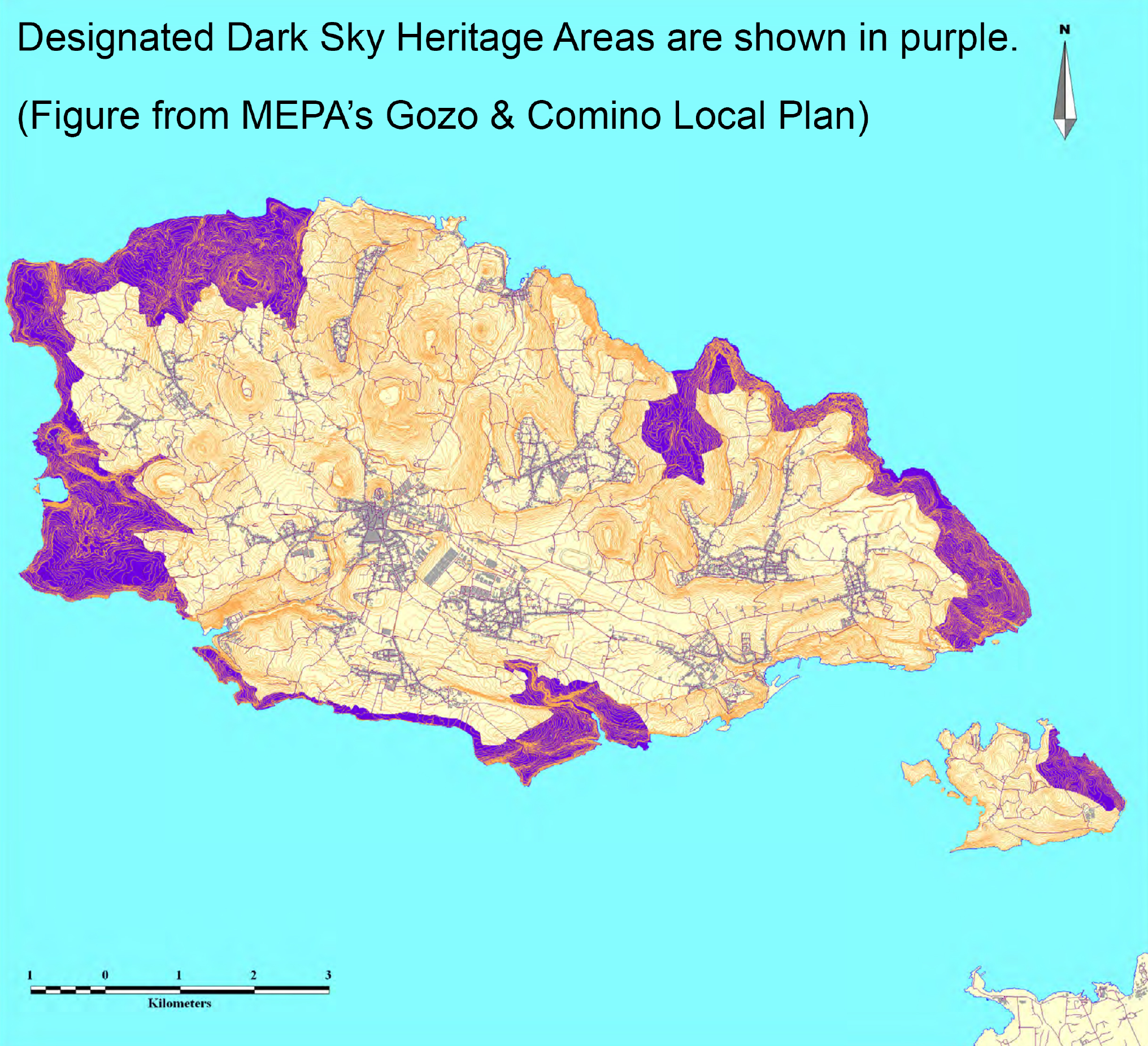}
  \caption{This figure reproduced from the Gozo and Comino Local Plan \citep{localplan} shows coastal zones that were designated as Dark Sky Heritage Areas in purple.}
  \label{fig:sf_01}
\end{figure}

\begin{figure}
  \includegraphics[width=\linewidth]{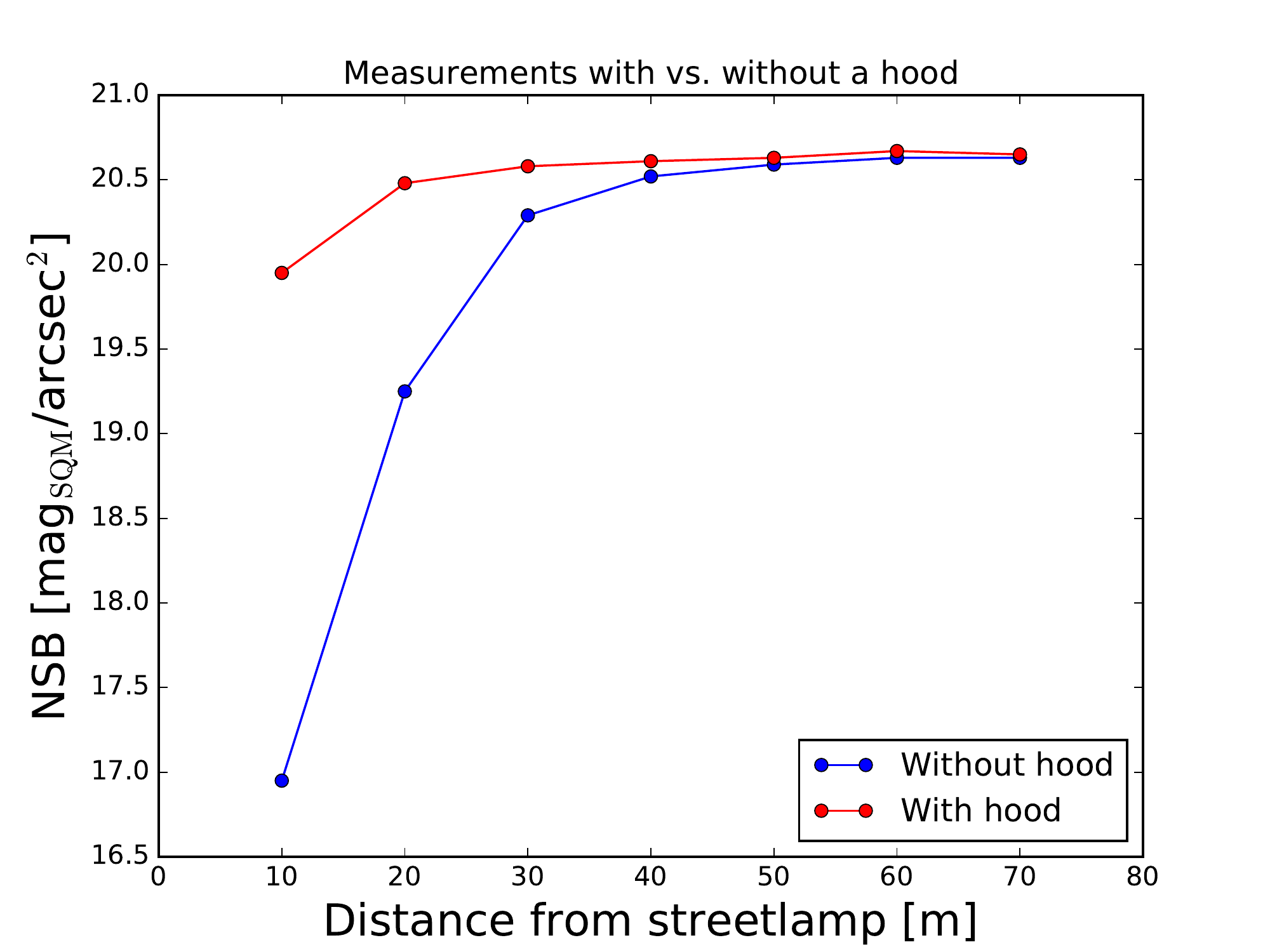}
  \caption{This figure shows the results from tests carried out to test the efficacy of a hood installed on the Sky Quality Meter.  At a distance of 10 m away from a streetlamp, the measurement recorded by the meter without a hood installed ($\approx 17$ mag$_{\rm SQM}$/arcsec$^2$) is very different from that registered when using a hood ($\approx 20$ mag$_{\rm SQM}$/arcsec$^2$), with the values approaching each other at a distance of 40 m.}
  \label{fig:sf_02}
\end{figure}

\begin{figure}
  \includegraphics[width=\linewidth]{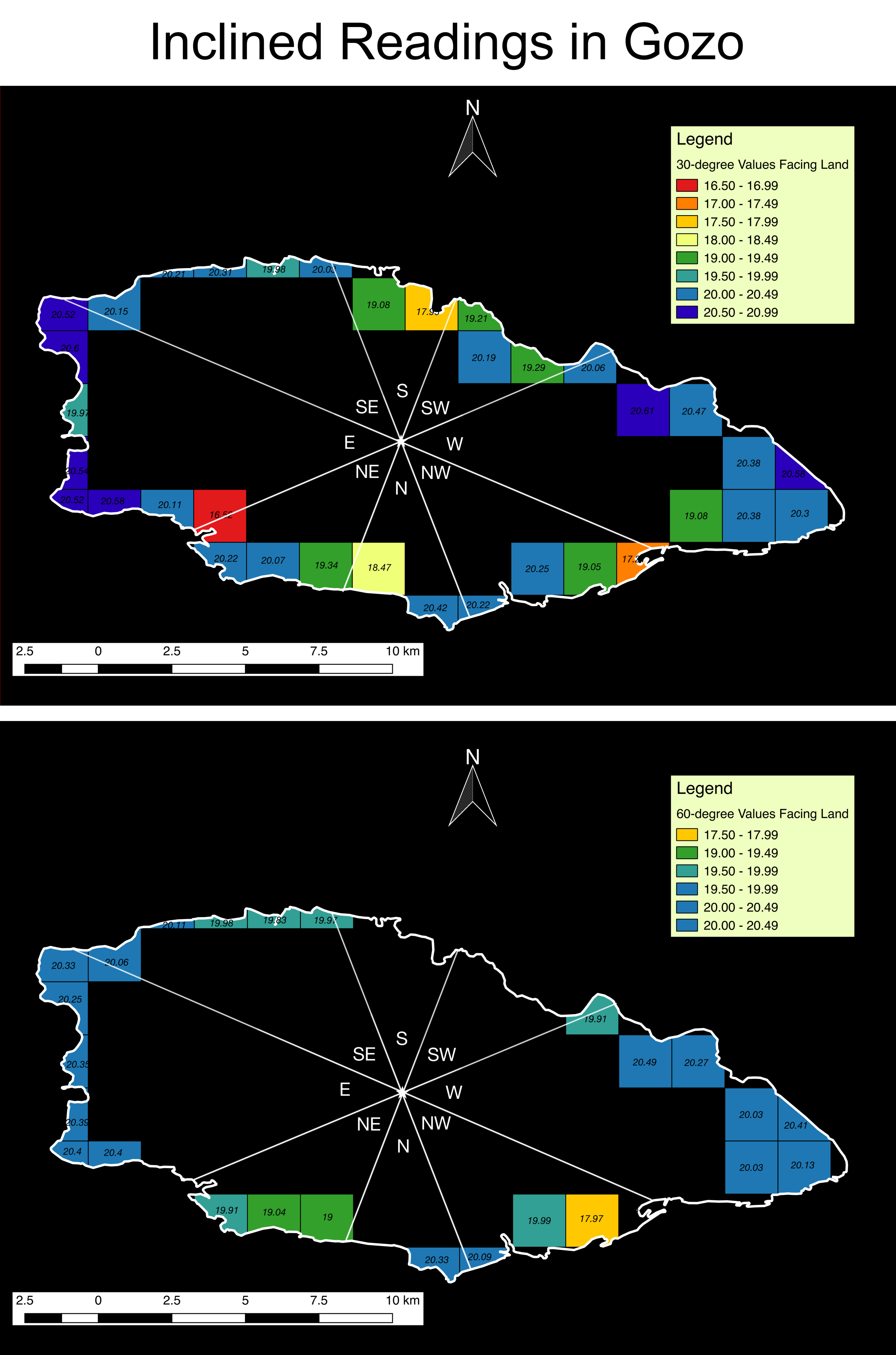}
  \caption{Inland-facing readings for Gozo at a zenith angle of 30$^{\rm{o}}$ (top panel) and 60$^{\rm{o}}$ (bottom panel), which show the brightness level that would be perceived when looking towards land, such as is the case for Shearwaters approaching the Island's cliffs.}
  \label{fig:sf_03}
\end{figure}

\begin{figure}
  \includegraphics[width=\linewidth]{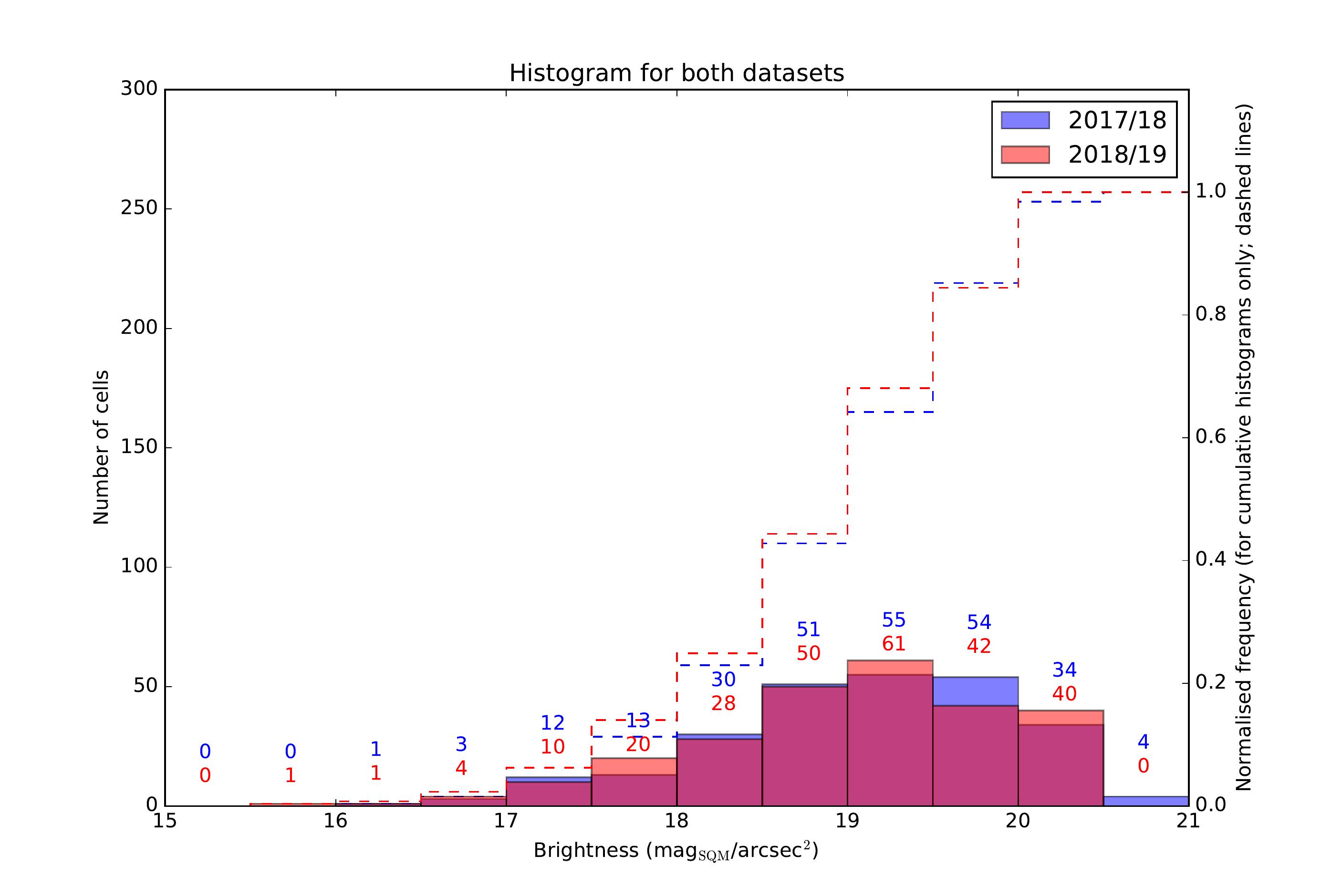}
  \caption{Frequency histograms of the NSB for the island of Malta for both of our geographical datasets, with filled areas denoting number counts, and cumulative histograms (dashed lines) overlaid.  The peak for both datasets occurs at $19-19.5$ mag$_{\rm SQM}$/arcsec$^2$.}
  \label{fig:sf_04}
\end{figure}

\begin{figure}
  \includegraphics[width=\linewidth]{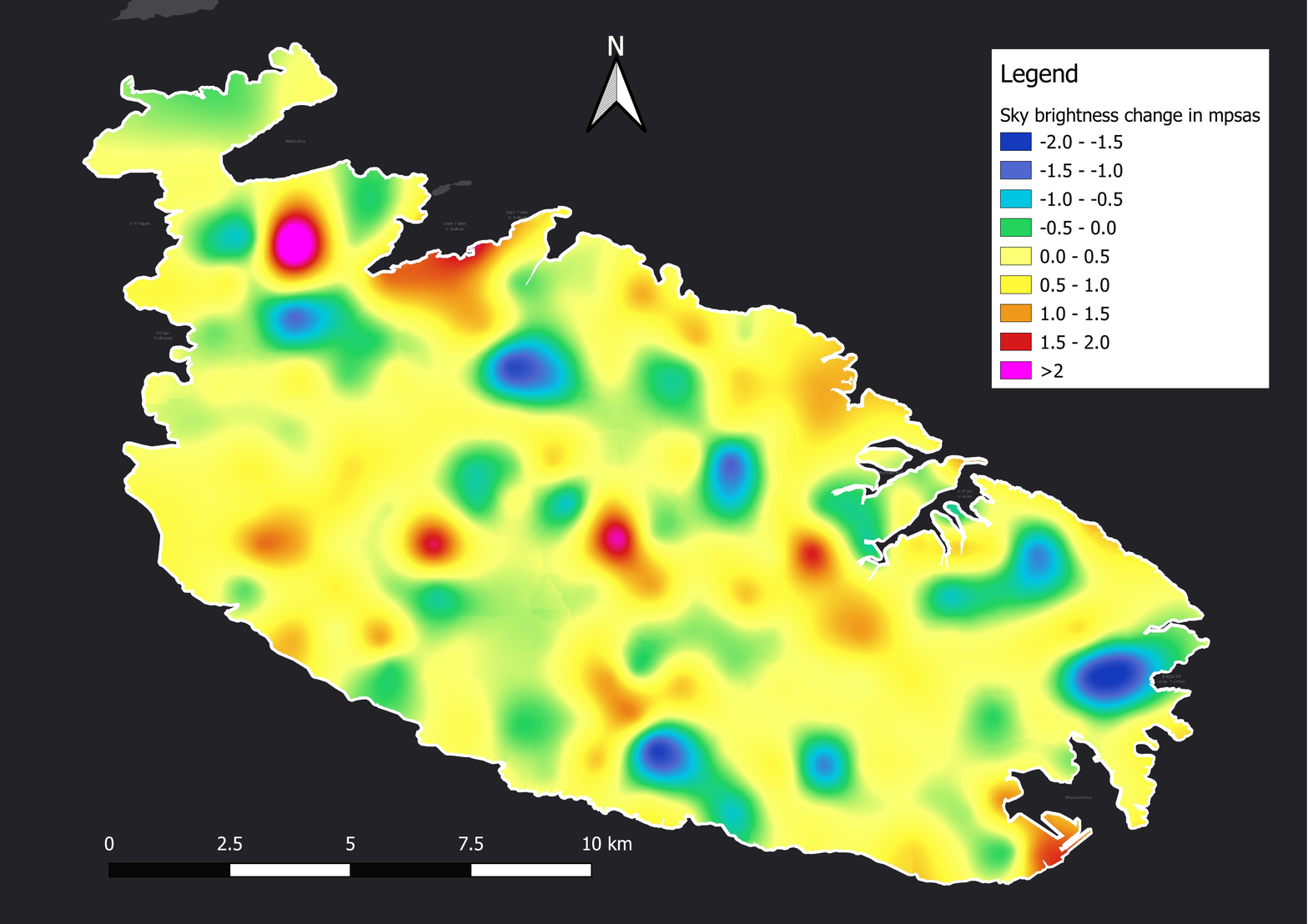}
  \caption{TPS difference map for the island of Malta that shows the change in NSB between the two datasets.  In the legend, a positive (negative) value denotes an increase (decrease) in NSB between 2017/18 and 2018/19.}
  \label{fig:sf_05}
\end{figure}

\begin{figure}
  \includegraphics[width=\linewidth]{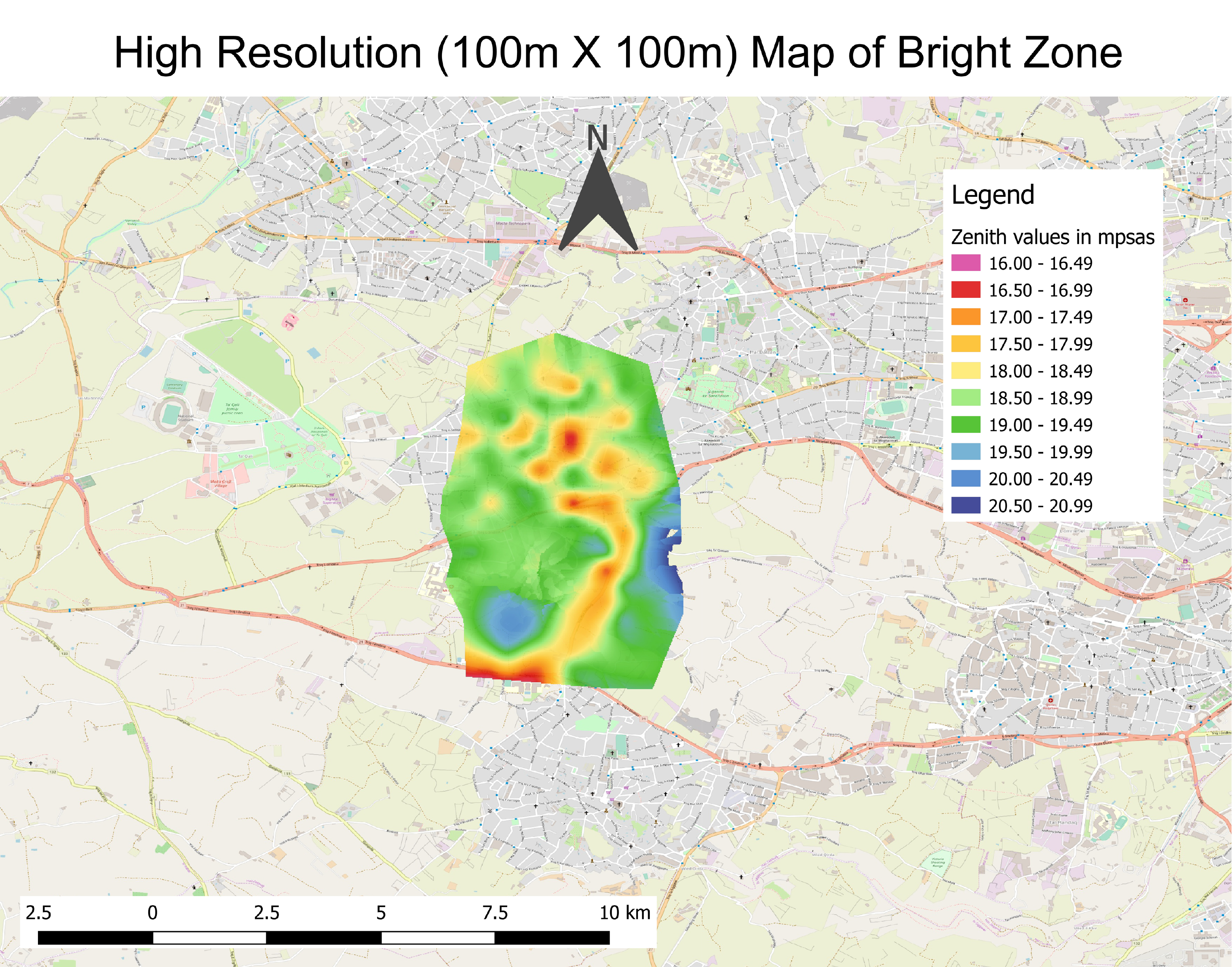}
  \caption{A high-resolution (100m$\times$100m) map of the area of Attard allows us to resolve structures that contribute to increased NSB, e.g.~showroom lighting.}
  \label{fig:sf_06}
\end{figure}

\begin{figure}
  \includegraphics[width=\linewidth]{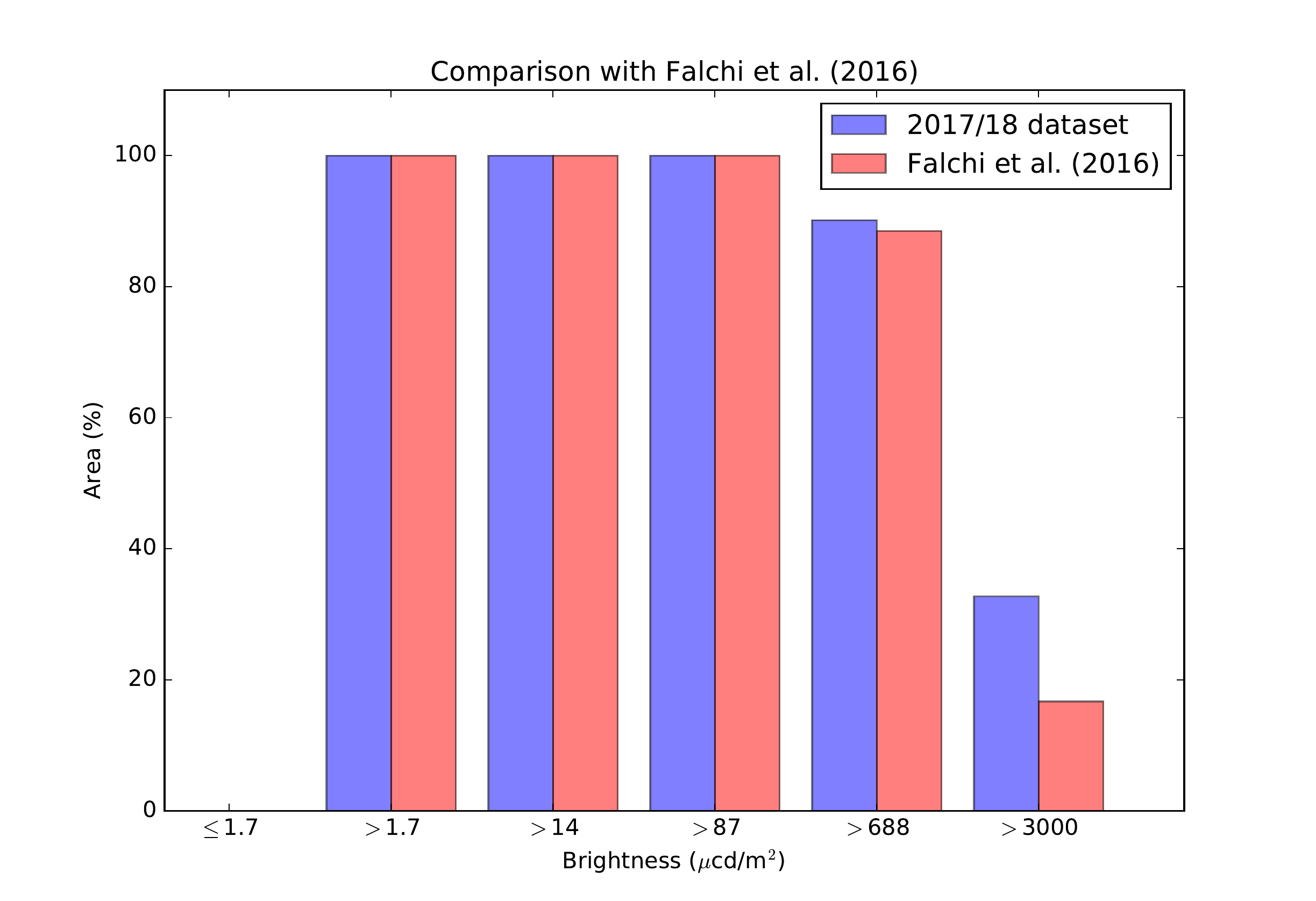}
  \caption{A bar chart for the Maltese archipelago showing the percentage area exhibiting a NSB larger than a given threshold.  Blue bars denote our measurements, basing upon the 2017/18 dataset, whereas red bars denote values predicted by \citet{falchi2016} - see their Table 2.}
  \label{fig:sf_07}
\end{figure}

\begin{figure}
  \includegraphics[width=\linewidth]{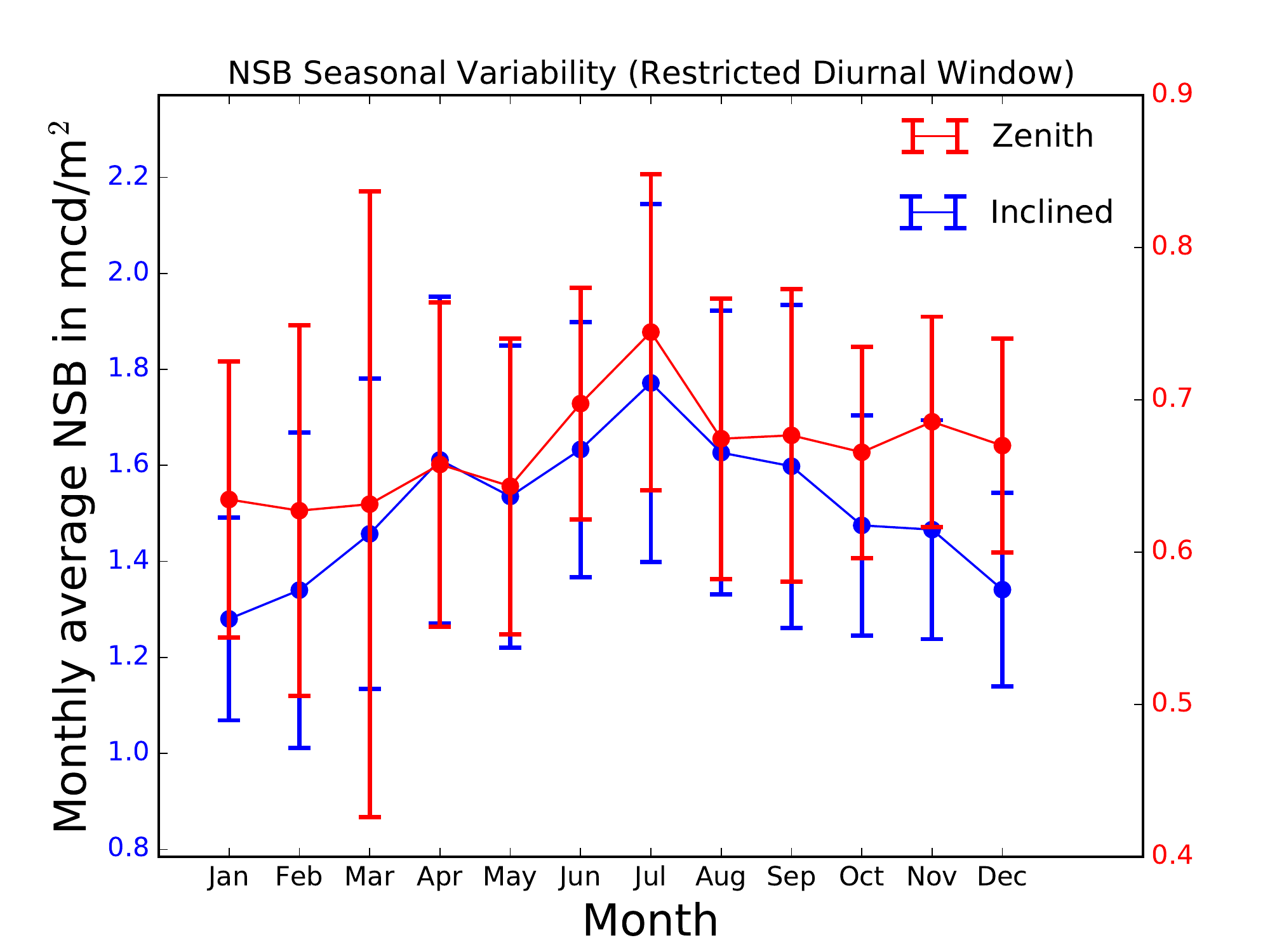}
  \caption{This figure shows the longterm NSB measured between 2014 and 2019 from a fixed site in Gozo.  It is similar to Fig.~\ref{fig:seasonalvariability} but restricts the diurnal window to between 22:10 and 03:55.  The same seasonal pattern is observed, exhibiting an increase in the NSB over the summer months.}
  \label{fig:sf_08}
\end{figure}

\begin{figure}
  \includegraphics[width=\linewidth]{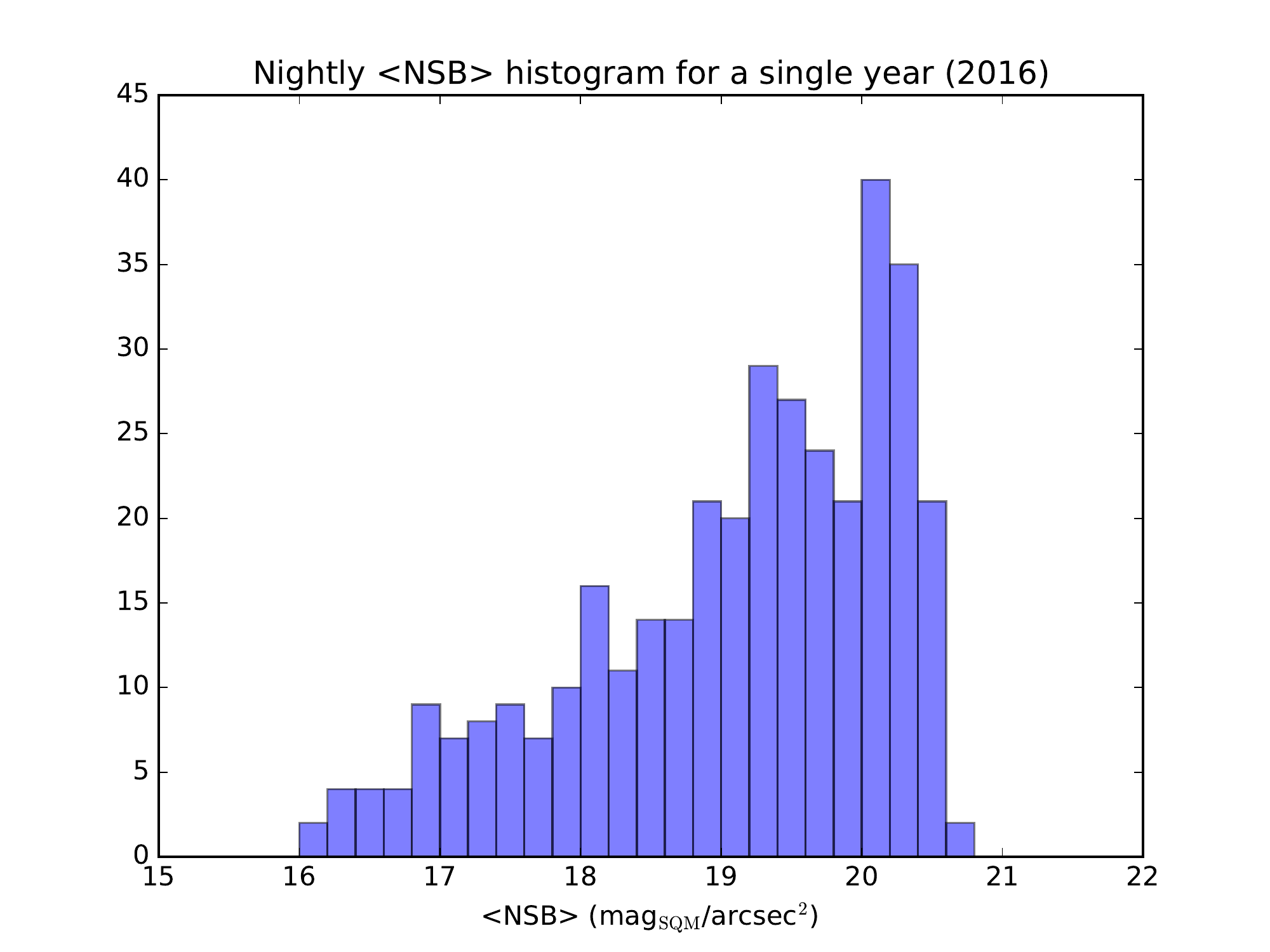}
  \caption{This histogram shows the (astronomical) nightly mean NSB, denoted as $\langle$NSB$\rangle$, for a single year (2016).  This plot considers all the data for that year, including cloudy and moonlit nights.  A peak for clear, moonless conditions is observed at $20-20.2$ mag$_{\rm SQM}$/arcsec$^2$; for comparison, measurements in Eastern Austria carried out during the same year by \citet{posch2018} yield a peak at 21.6 mag$_{\rm SQM}$/arcsec$^2$.}
  \label{fig:sf_09}
\end{figure}

\begin{figure}
  \includegraphics[width=\linewidth]{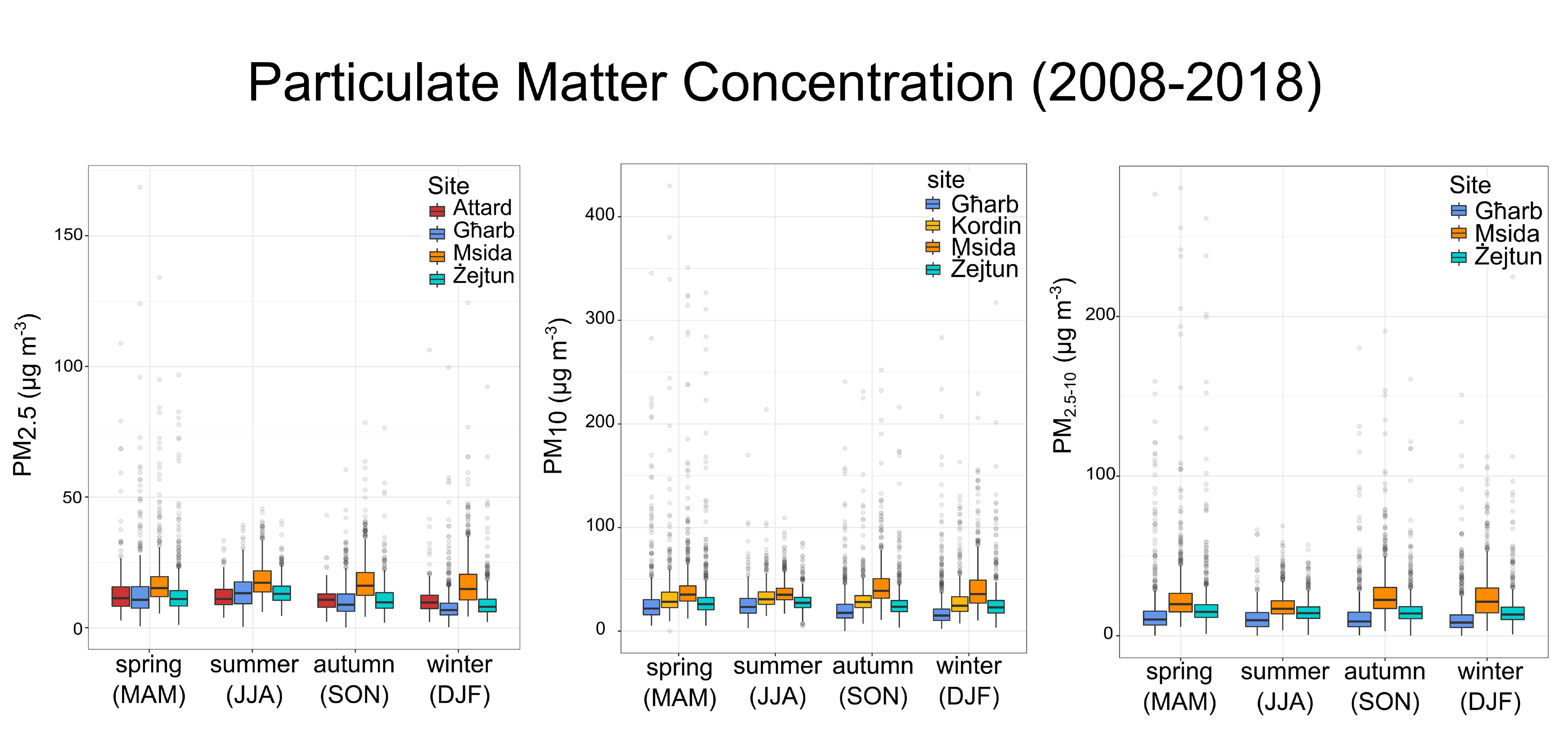}
  \caption{This figure shows seasonal particulate matter (PM) concentrations covering the period 2008-2018 obtained from the air quality network of the Environment and Resources Authority.  The left panel is for PM$_{2.5}$ (fine), the middle panel for PM$_{10}$ and the right panel for PM$_{\rm coarse}$ (calculated as PM$_{10}$ - PM$_{2.5}$).  In all three plots, PM concentrations are higher at the traffic site of Msida and lowest at the rural and urban background sites of G\malteseh arb and \.{Z}ejtun, respectively. The seasonality in PM concentrations is weak for all stations considered.  However, PM concentrations are marginally higher in autumn and winter at the Msida station compared to the other seasons, but lower in G\malteseh arb for the same seasons.  Boxes denote the interquartile range with the median being represented by a horizontal bar inside.  The upper (lower) error bar denotes the largest value that still lies within a factor of 1.5 of the interquartile range above the 75th (below the 25th) percentile, with values outside the error bars being outliers.}
  \label{fig:sf_10}
\end{figure}

\begin{figure}
  \includegraphics[width=\linewidth]{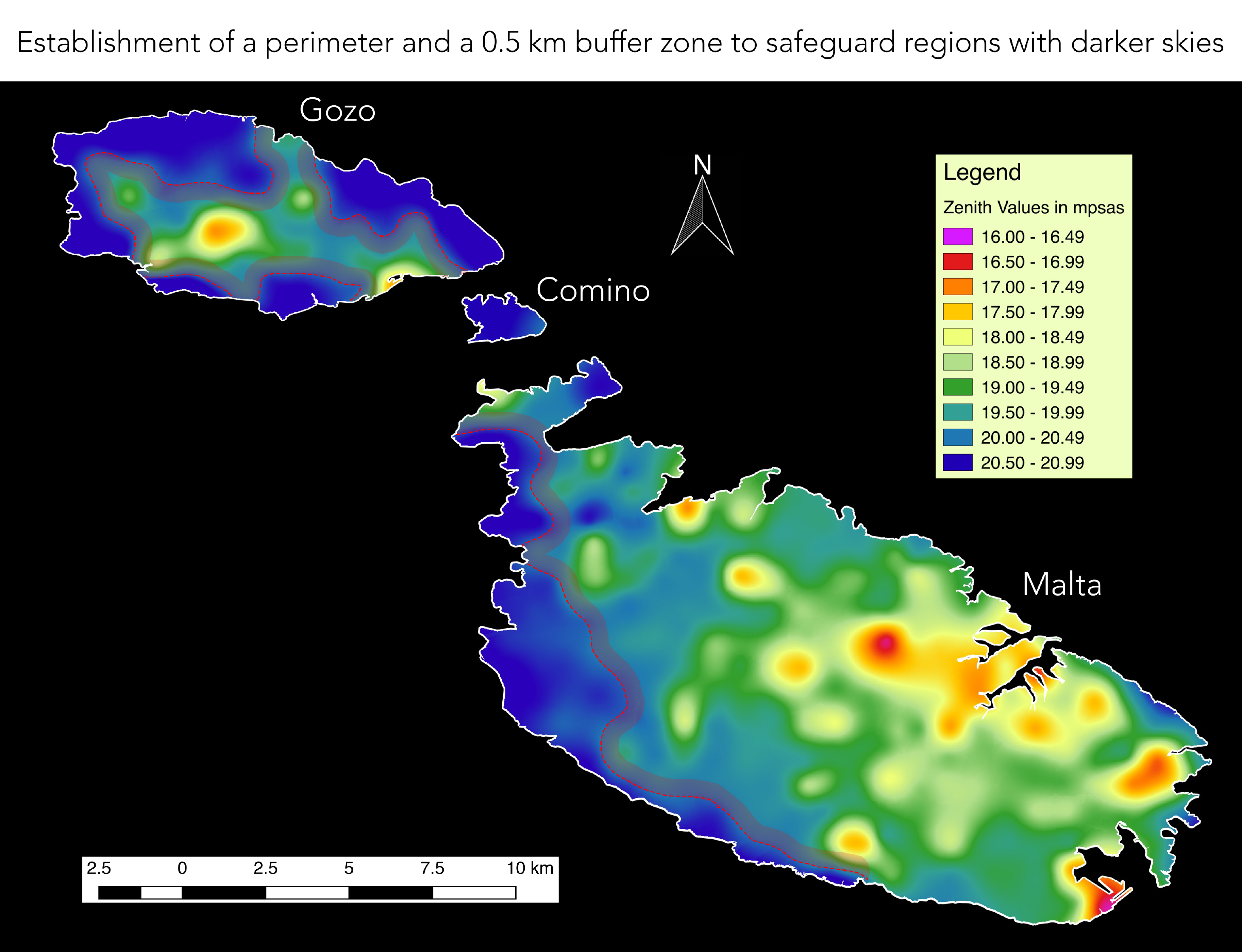}
  \caption{A proposed perimeter (dotted red line) and 0.5 km buffer zone (transparent red band) to protect darker zones on the two main islands of Malta and Gozo.}
  \label{fig:sf_11}
\end{figure}


\begin{thebibliography}{9}
\bibitem[Bara(2018)]{bara2018}
Bar\'{a}, S. Characterizing the zenithal night sky brightness in large territories: how many samples per square kilometre are needed? Mon.~Not.~R.~Astron.~Soc., 473 (3), 4164-4173.
\end{thebibliography}
\end{document}